\documentclass[11pt,a4paper]{article}
\usepackage{jcappub}
\usepackage[T1]{fontenc}
\usepackage{graphicx}
\usepackage{bm}
\usepackage{amsmath,amsfonts,amssymb}
\usepackage{mathtools}
\usepackage{braket}
\usepackage{setspace} 
\usepackage{here}
\usepackage{xcolor}

\usepackage{soul}
\usepackage[normalem]{ulem}
\usepackage{cancel}

\begin{document}


\title{Resonant production of sterile neutrino dark matter with a refined numerical scheme}
\author[a]{Kentaro Kasai,}
\author[a,b]{Masahiro Kawasaki,}
\author[c]{and Kai Murai}
\affiliation[a]{ICRR, University of Tokyo, Kashiwa 277-8582, Japan}
\affiliation[b]{Kavli IPMU (WPI), UTIAS, University of Tokyo, Kashiwa 277-8583, Japan}
\affiliation[c]{Department of Physics, Tohoku University, Sendai 980-8578, Japan}

\abstract{
The existence of a large primordial neutrino asymmetry is an intriguing possibility, both observationally and theoretically. Such an asymmetry can lead to the resonant production of $\mathrm{keV}$-scale sterile neutrinos, which are a fascinating candidate for dark matter. In this paper, we comprehensively revisit the resonant production processes with a refined numerical analysis, adopting a dynamical discretization of momentum modes to take care of the sharpness of the resonance.
We find parameter regions consistent with X-ray and Lyman-$\alpha$ constraints for lepton-to-entropy ratio $\gtrsim \mathcal{O}(10^{-3})$ and $m_{\nu_s}\gtrsim 20$\,keV. 
We also explore the Affleck-Dine mechanism as a possible origin for such asymmetries. 
While previous studies considered resonant production after lepton number generation, we numerically investigate cases where a fraction of sterile neutrinos is produced during lepton number injection. 
In this regime, some parameter sets can shorten the free-streaming length and reduce the required mixing angle to match the observed dark matter abundance, thereby mitigating the observational constraints.
}


\emailAdd{kkasai@icrr.u-tokyo.ac.jp}
\emailAdd{kawasaki@icrr.u-tokyo.ac.jp}
\emailAdd{kai.murai.e2@tohoku.ac.jp}

\begin{flushright}
    TU-1279
\end{flushright}
\maketitle

\section{Introduction}
\label{sec: intro}

Sterile neutrinos with masses of the $\rm{keV}$ scale are intriguing candidates for dark matter (DM) (see Refs.~\cite{Abazajian:2017tcc,Boyarsky:2018tvu, Shakya:2015xnx} for reviews).
They can be produced through neutrino oscillations between sterile neutrinos and the Standard Model (SM) neutrinos (``active'' neutrinos)~\cite{Dodelson:1993je,Shi:1998km,Abazajian:2001nj,Abazajian:2001vt} via mixing.

The simplest setup for this production is known as the Dodelson-Widrow (DW) mechanism~\cite{Dodelson:1993je, Abazajian:2005gj}.
However, the vacuum mixing angle required to realize the correct DM abundance and the resultant free-streaming length of sterile neutrinos are severely constrained by current observational data. 
In particular, the X-ray observations of galaxies, which are sensitive to the radiative decay of sterile neutrinos~\cite{Shrock:1974nd,Dolgov:2000ew,Abazajian:2001vt,Boyarsky:2005us,Boyarsky:2006fg,denHerder:2009sxr}, and the structure formation constraints from Lyman-$\alpha$ forest data~\cite{Viel:2005qj,Baur:2017stq} already exclude the DW scenario at more than 2$\sigma$ level.

These stringent constraints can be significantly relaxed if there exists a large primordial neutrino-anti-neutrino asymmetry with the lepton-to-entropy ratio $\gtrsim \mathcal{O}(10^{-5})$~\cite{Abazajian:2001nj,Abazajian:2001vt,Kishimoto:2008ic,Laine:2008pg,Boyarsky:2009ix, Ghiglieri:2015jua, Venumadhav:2015pla,Gelmini:2019clw, Chen:2025sgd}. 
In such a case, the non-zero lepton number modifies the self-energy of active neutrinos, leading to a resonant enhancement of the production rate at a certain cosmic temperature in the early universe.
Consequently, we can explain all DM with a smaller vacuum mixing angle than in the DW mechanism, satisfying the X-ray constraints. 
This possibility is also motivated by the recently reported primordial abundance of helium-4 inferred from observation of extremely metal-poor galaxies~\cite{Matsumoto:2022tlr, Yanagisawa:2025mgx}, which suggests the existence of a large primordial electron neutrino-anti-neutrino asymmetry before big bang nucleosynthesis (BBN).

This paper aims to provide a comprehensive scan of the model parameters that can explain all DM density by resonantly produced sterile neutrinos consistent with X-ray observations. 
Although several previous works have studied the same subject, the early studies~\cite{Laine:2008pg, Boyarsky:2009ix} assume that the lepton asymmetry in SM neutrinos is always equilibrated among the flavors in the early universe. 
However, various studies~\cite{Dolgov:2002ab, Mangano:2010ei, Mangano:2011ip, Barenboim:2016shh, Johns:2016enc, Froustey:2024mgf} have shown that flavor oscillations within SM neutrinos take place well after the production of the sterile neutrinos.
Thus, the assumption of equilibration would be invalid.
More recent studies~\cite{Ghiglieri:2015jua, Venumadhav:2015pla} have modified these assumptions and performed more realistic analyses, although they do not offer a comprehensive analysis of the parameter space. 
Moreover, some numerical difficulties remain due to the sharpness of the resonance unless the momentum discretization is treated with sufficient care, especially when the initial lepton asymmetry is large and/or the mass of the sterile neutrino is small~\cite{Kasai:2024diy, Akita:2025txo}. 

In this paper, we reexamine the resonant production scenario with careful numerical calculations of the momentum distribution of sterile neutrinos near the resonance and discuss the observational constraints. 
We find that the lepton-to-entropy ratio is required to be $\gtrsim \mathcal{O}(10^{-3})$ to evade both the X-ray and Lyman-$\alpha$ constraints. 
We also investigate the sterile neutrino production when the lepton asymmetry of each flavor is large but the total asymmetry vanishes, which is suggested by the very recent papers~\cite{Vogel:2025aut,Akita:2025txo}.
In this case, the flavor oscillations can reduce the asymmetries after the sterile neutrinos production and before the onset of BBN, which significantly relaxes the constraints on the lepton asymmetry from BBN.

Furthermore, we discuss the generation of such a large lepton asymmetry within the framework of Affleck-Dine leptogenesis~\cite{Affleck:1984fy,Dine:1995kz,Kawasaki:2002hq,Kawasaki:2022hvx,Kasai:2024diy, Akita:2025zvq}. 
We investigate the case in which non-topological solitons called Q-balls are formed by a leptonic scalar field. 
Lepton asymmetry stored in Q-balls is protected from the sphaleron process and is released as active neutrinos when the Q-balls decay well after the electroweak phase transition, which successfully explains the origin of a large neutrino asymmetry without generating too large baryon asymmetry. 
In the context of resonantly produced sterile neutrino DM, the main idea was discussed in our previous paper~\cite{Kasai:2024diy}. 
However, just for simplicity, we only considered the resonant production after the Q-balls completely decay.
In contrast, in this paper, we show that the final spectrum is significantly altered if a substantial fraction of sterile neutrinos is produced during the decay of Q-balls. 
Notably, if sterile neutrinos are produced around the time when the decay process is completed, it can relax the observational constraints from X-ray and structure formation.

The rest of this paper is organized as follows. 
In Sec.~\ref{sec : resonant production}, we outline the basic formalism for the resonant production of sterile neutrinos.
In Sec.~\ref{sec : numerical setup}, we describe the numerical methods used to solve these equations, and present the results for the required mixing angle $\theta$ to match the DM abundance, along with the resulting spectrum, free-streaming length, and the observational constraints. 
In Sec.~\ref{sec : generation of a large lepton asymmetry}, we investigate a concrete leptogenesis scenario based on the Affleck-Dine mechanism. 
We also numerically investigate the cases where sterile neutrinos are produced during lepton number injection, and show the observational constraints. 
Finally, Sec.~\ref{sec : discussion and summary} provides the summary of our results. 

\section{A basic formalism for resonant production of sterile neutrinos}
\label{sec : resonant production}

In this section, we briefly summarize the basic equations that describe the evolution of sterile neutrinos and lepton asymmetry. 

\subsection{Basic equations}

In this paper, we consider the mixing between one flavor of the SM neutrinos and one sterile neutrino species as
\begin{equation}
\begin{aligned}
    \ket{\nu_a}
    &=\cos\theta\ket{\nu_1}
    +\sin\theta\ket{\nu_2},
    \\
    \ket{\nu_s}
    &=-\sin\theta\ket{\nu_1}
    +\cos\theta\ket{\nu_2},
\end{aligned}
\end{equation}
where $\ket{\nu_a}$ and $\ket{\nu_s}$ are flavor eigenstates of an active neutrino ($a = e$, $\mu$, or $\tau$) and a sterile neutrino, respectively, $\ket{\nu_1}$ and $\ket{\nu_2}$ are the mass eigenstates, and $\theta$ is their mixing angle in vacuum.
The mixing of the anti-particle states is also described in the same way. 
While we explain the formalism for general $a$ here, we will mainly discuss the case of $a = e$ in the subsequent sections. 
We also introduce a density matrix involving the two flavor eigenstates as
\begin{equation}
\begin{aligned}
    \rho (t, p)
    &=
    \begin{pmatrix}
        \rho_{aa} & \rho_{as} \\
        \rho_{as}^* & \rho_{ss} \\
    \end{pmatrix}
    \ ,
    \\
    \bar{\rho} (t, p)
    &=
    \begin{pmatrix}
        \bar{\rho}_{aa} & \bar{\rho}_{as} \\
        \bar{\rho}_{as}^* & \bar{\rho}_{ss} \\
    \end{pmatrix}
    \ , 
\end{aligned}
\end{equation}
where $\rho_{aa}$ and $\rho_{ss}$ represent the momentum distribution functions of the active and sterile neutrinos, respectively, and $\rho_{as}$ represents the coherence between the two states. 
$\bar{\rho}$ represents the same stuff of the anti-particles. 

The effective description of the kinetic equation of neutrino states is given by~\cite{Dolgov:2002wy,Shaposhnikov:2008pf}
\begin{equation}
    i(\partial_t-Hp\partial_p) \rho
    =
    \left[ \mathcal{H},\rho \right]
    -i\{ \Gamma,\rho-\rho_{\rm{eq}} \}
    \ ,
    \label{eq : density operator equation}
\end{equation}
where $H$ is the Hubble parameter, $p$ is the physical momentum, $\mathcal{H}$ is the Hamiltonian, and $[\cdot\, ,\, \cdot]$ and $\{\cdot\, , \,\cdot\}$ denote 
commutator and anti-commutator, respectively.
Here, $\rho_{\rm{eq}}$ is the thermal distribution given by
\begin{equation}
    \rho_{\rm{eq}}
    =
    \begin{pmatrix}
        f_{\rm{eq}}(\sqrt{m_{\nu_a}^2+p^2},\mu_{\nu_a}) & 0 \\
        0 & f_{\rm{eq}}(\sqrt{m_{\nu_s}^2+p^2},\mu_{\nu_s}) \\
    \end{pmatrix}
    \ ,
    \label{eq : density matrix describing thermal distribution}
\end{equation}
where $m_{\nu_a}$ and $m_{\nu_s}$ are the active and sterile neutrino masses, respectively, $f_{\rm{eq}}(E,\mu) \equiv 1/(\exp[(E-\mu)/T] + 1)$ is the Fermi-Dirac distribution function, and $\mu_{\nu_a}$ and $\mu_{\nu_s}$ are the chemical potentials of the active and sterile neutrinos, respectively. 
Here, we use the mass eigenvalues of $\ket{\nu_1}$ and $\ket{\nu_2}$ for $m_{\nu_a}$ and $m_{\nu_s}$, respectively, which can be justified in the limit of a small mixing angle.
In the following, we approximate that $\sqrt{m_{\nu_a}^2+p^2}\simeq\sqrt{m_{\nu_s}^2+p^2}\simeq p$ since both the active and sterile neutrinos are relativistic at the epoch of interest.
$\Gamma$ is a Hermitian matrix associated with the scattering rate of neutrinos in the thermal plasma.
Since sterile neutrinos do not interact with the thermal plasma, $\Gamma$ has a form of
\begin{equation}
    \Gamma
    =
    \frac{1}{2}
    \begin{pmatrix}
        \Gamma_{\nu_a} & 0 \\
               0 & 0 \\
    \end{pmatrix}
    \ ,
    \label{eq : matrix describing scattering}
\end{equation}
where $\Gamma_{\nu_a}$ is the total thermal width of the active neutrino given by
\begin{align}
    \Gamma_{\nu_a} (p,T)
    &=
    y_a(p,T) G_{\rm{F}}^2 p T^4 
    \ .
    \label{eq : thermal width}
\end{align}
Here, $G_{\rm{F}}$ is the Fermi coupling constant, and $y_a(p,T)$ is a factor depending on the degrees of freedom in the thermal plasma.
See Appendix~\ref{append : ye fitting} for the fitting formula of $y_e$ used in our numerical calculations.
In the flavor basis, the Hamiltonian $\mathcal{H}$ is written as
\begin{equation}
    \mathcal{H}
    =
    \begin{pmatrix}
        V_a-\frac{m_{\nu_s}^2}{4p}\cos2\theta & \frac{m_{\nu_s}^2}{4p}\sin2\theta 
        \vspace{5pt} \\
        \frac{m_{\nu_s}^2}{4p}\sin2\theta & \frac{m_{\nu_s}^2}{4p}\cos2\theta
        \end{pmatrix}
        \ ,
    \label{eq : Hamiltonian matrix}
\end{equation}
where $V_a$ denotes the effective potential due to the finite temperature and density effects on the self-energy of the active neutrino.
The first term in the right-hand side of Eq.~\eqref{eq : density operator equation} describes the neutrino oscillation effect, and the second term describes the collision process of active neutrinos.
Consequently, Eq.~\eqref{eq : density operator equation} can be rewritten as
\begin{align}
    i(\partial_t-Hp\partial_p)\rho_{aa}
    &=
    \frac{m_{\nu_s}^2}{4p}\sin2\theta(\rho_{as}^*-\rho_{as})-i\Gamma_{\nu_a}(\rho_{aa}-f_{\rm{eq}}(p,\mu_{\nu_a}))
    \ , 
    \label{eq : density operator equation of rho aa}
    \\
    i(\partial_t-Hp\partial_p)\rho_{ss} 
    &=
    \frac{m_{\nu_s}^2}{4p}\sin2\theta(\rho_{as}-\rho_{as}^*)
    \ , 
    \label{eq : first line in density operator equation}
    \\
    i(\partial_t-Hp\partial_p)\rho_{as} 
    &=
    \frac{m_{\nu_s}^2}{4p}\sin2\theta(\rho_{ss}-\rho_{aa})
    +
    \left(
        V_a -\frac{m_{\nu_s}^2}{2p}\cos2\theta
        -i\frac{\Gamma_{\nu_a}}{2}
    \right) \rho_{as}
    \ .
    \label{eq : second line in density operator equation}
    \end{align}
Since $\Gamma_{\nu_a}$ is much larger than $H$ and $m_{\nu_s}^2/p$ for the temperatures of interest, Eq.~\eqref{eq : density operator equation of rho aa} leads to $\rho_{aa} \simeq f_\mathrm{eq}(p,\mu_{\nu_a})$.

The thermal self-energies of the active and anti-active neutrinos, $V_a$ and $V_{\bar{a}}$, are respectively given by~\cite{Notzold:1987ik}
\begin{equation}
\begin{aligned}
    V_a 
    &=
    \sqrt{2}G_{\rm{F}} 
    \left(2(n_{\nu_a}-n_{\bar{\nu}_a})
    + \sum_{b \neq a}(n_{\nu_b}-n_{\bar{\nu}_b})
    +
    \left(
    \frac{1}{2}+2\sin^2\theta_{w}
    \right)
    (n_{e_a}-n_{\bar{e}_a})\right.
    \\
    &\left. \qquad \qquad \quad  -\left(
    \frac{1}{2}-2\sin^2\theta_{w}
    \right)
    \sum_{b \neq a}
    (n_{e_b}-n_{\bar{e}_b})
    \right)
    -B_ap T^4
    \\
    &\equiv  \sqrt{2}G_{\rm{F}}s_{\rm{tot}}
    \mathcal{L}_a-B_apT^4,
    \\
    V_{\bar{a}} 
    & =
    -\sqrt{2}G_{\rm{F}}s_{\rm{tot}}
    \mathcal{L}_a-B_apT^4,
    \label{eq : finite density potential}
\end{aligned}
\end{equation}
where $n_{\nu_a}$ and $n_{\bar{\nu}_a}$ are the number densities of 
$\nu_a$ and $\bar{\nu}_a$, respectively, $n_{e_a}$ and $n_{\bar{e}_a}$ are the number densities of the charged leptons and antileptons, respectively, $\theta_w$ is the Weinberg angle satisfying $\sin^2\theta_w\simeq 0.23$, and $s_{\rm{tot}}$ is the total entropy density.
Here, the subscripts $a,b\, (=e,\mu,\tau)$ denote the flavors of the active neutrinos, and $B_a$ is a constant taking a value of $B_e=10.88 \times 10^{-9}\,{\rm{GeV}}^{-4}$ for $a=e$~\cite{Gelmini:2019wfp}. Furthermore, we have defined a potential lepton number $\mathcal{L}_{a}$ as 
\begin{align}
    \mathcal{L}_a
    \equiv &
    \frac{1}{s_\mathrm{tot}}
    \left[2(n_{\nu_a}-n_{\bar{\nu}_a})
    + \sum_{b \neq a}(n_{\nu_b}-n_{\bar{\nu}_b})
    +
    \left(
    \frac{1}{2}+2\sin^2\theta_{w}
    \right)
    (n_{e_a}-n_{\bar{e}_a})
    \right. 
    \nonumber \\
    & \qquad \left. -\left(
    \frac{1}{2}-2\sin^2\theta_{w}
    \right)
    \sum_{b \neq a}
    (n_{e_b}-n_{\bar{e}_b})\right].
\label{eq: def of potential L}
\end{align}
In Eq.~\eqref{eq : finite density potential}, we neglected the contributions of the baryon sector because they are much smaller than those of the active neutrino sector as discussed in Refs.~\cite{Laine:2008pg,Ghiglieri:2015jua}.

If there is no lepton asymmetry, i.e., $\mathcal{L}_a = 0$,  $V_a$ and $V_{\bar{a}}$ are equal and negative ($V_a = V_{\bar{a}} < 0$), and hence $\nu_s$ and $\bar{\nu}_s$ are produced with the same amount via the DW mechanism.
On the other hand, if $\mathcal{L}_a$ is positive and large enough, $V_a$ becomes positive, and the two diagonal components of the Hamiltonian in Eq.~\eqref{eq : Hamiltonian matrix} become equal at some temperature, which indicates a maximal mixing of the active and sterile neutrinos. 
Consequently, $\nu_s$ is resonantly produced while the production of $\bar{\nu}_s$ is not so different from the DW mechanism. 
The condition for the resonance is given by
\begin{align}
    V_a-\frac{m_{\nu_s}^2}{4p}\cos2\theta&=\frac{m_{\nu_s}^2}{4p}\cos2\theta
    \nonumber \\
    \Leftrightarrow \quad 
    1-\frac{2p}{m_{\nu_s}^2}V_a & \simeq 
    0,
    \label{eq : resonance condition 1}
\end{align}
where we used $\cos2\theta\simeq 1$.
If $\mathcal{L}_a < 0$, $\bar{\nu}_{s}$ can be resonantly produced in a similar way.

\subsection{Time Evolution of lepton asymmetry}

Here, we describe the time evolution of lepton asymmetry and the asymmetry redistribution among the active neutrinos and charged leptons, basically following Refs.~\cite{Venumadhav:2015pla, Ghiglieri:2015jua}. First of all, we define a lepton asymmetry $L_a$ of an active flavor $a=e,\mu,\tau$ by
\begin{align}
    L_a
    \equiv
    \frac{
    n_{\nu_a}-n_{\bar{\nu}_a}
    +
    n_{e_a}-n_{\bar{e}_a}
    }
    {s_{\rm{tot}}}.
    \label{eq : def of La}
\end{align}
Now, since the sterile neutrino production occurs at $T=\mathcal{O}(0.1)$\,GeV as we will see later, the flavor oscillations do not take place around or before the production of the sterile neutrinos~\cite{Dolgov:2002ab, Mangano:2010ei, Mangano:2011ip, Barenboim:2016shh, Johns:2016enc, Froustey:2024mgf}.
Thus, we consider that $L_a$ of flavors that do not mix with $\nu_s$ is conserved. On the other hand, regarding the flavor mixed with $\nu_s$, we impose the condition
\begin{align}
    L_{a}+\frac{n_{\nu_s}-n_{\bar{\nu}_s}}{s_{\rm{tot}}}
    =
    {\rm{const}}.
\end{align}
Then, we obtain
\begin{align}
\frac{{\rm{d}}}{{\rm{d}}t}
L_{a}
&= -\int \frac{p^2{\rm{d}}p}{2\pi^2}
\frac{(\partial_t-Hp\partial_p)
(\rho_{ss}-\bar{\rho}_{ss})}{s_{\rm{tot}}}=
-\frac{m_{\nu_s}^2\sin2\theta}
{s_{\rm{tot}}}\int \frac{p{\rm{d}}p}{4\pi^2}
\;{\rm{Im}}(\rho_{as}-\bar{\rho}_{as}),
\label{eq : time evolution of L}
\end{align}
where we used Eq.~\eqref{eq : first line in density operator equation} and its counterpart for the anti-particle in the second equality.

Now, we discuss the redistribution of lepton asymmetry among neutrinos and charged leptons. At the epoch of $\nu_s$ production, the process of 
    $\nu_a+e_b 
    \leftrightarrow 
    \nu_b+e_a$ 
is in chemical equilibrium. 
Furthermore, when the cosmic temperature is higher than 
that of QCD phase transition, $T_{\rm{QCD}}$, the reaction  
    $\nu_a+\bar{e}_a
    \leftrightarrow
    \alpha+\bar{\beta}$,
where $\alpha$ and $\beta$ are quarks of up-type and down-type, respectively, is also in chemical equilibrium. 
On the other hand, at $T\lesssim T_{\rm{QCD}}$, the reactions 
    $\pi^+\leftrightarrow \nu_a+\bar{e}_a$ 
for $a=e,\mu$ are in chemical equilibrium. 
They redistribute a fraction of lepton asymmetry into charged leptons at the epoch of $\nu_s$ production.
This effect slightly modifies the final result because the effects of neutrino asymmetry and charged lepton asymmetry on $V_a$ are different, as shown in Eq.~\eqref{eq : finite density potential}. 
We can determine the charged-lepton asymmetry by requiring charge neutrality and neglecting baryon asymmetry, which is much smaller than the lepton asymmetry of interest.
To do this, we define chemical potentials with respect to the electromagnetic charge density, baryon number density, and lepton flavor density as $\mu_q$, $\mu_B$, and $\mu_a$, respectively.
Then, the electromagnetic charge density $n_q$, baryon number density $n_B$, and $L_a$ are given at linear order in chemical potentials by
\begin{align}
    n_q
    &=
    \chi_q(T)\mu_q+\chi_{Bq}(T)\mu_B 
    -2\sum_a\chi(m_{e_a},T)(\mu_a-\mu_q)
    =0,
    \nonumber 
    \\
    n_B&=
    \chi_B(T)\mu_B+\chi_{Bq}(T)\mu_q
    \simeq 0,
    \label{eq : nB and nQ and na}
    \\
    L_a
    &=
    \frac{
    \chi(0,T)\mu_a
    +
    2\chi(m_{e_a},T)
    (\mu_a-\mu_q)}{s_{\rm{tot}}},
    \nonumber
\end{align}
where $\chi(m,T)$, $\chi_B$, $\chi_q$, and $\chi_{Bq}$ are susceptibilities defined by
\begin{equation}
\begin{gathered}
    \chi(m,T)
    \equiv
    \int
    \frac{p^2{\rm{d}}p}{2\pi^2}
    \left.\frac{\partial}{\partial E}
    f_{\rm{eq}}(E,T)\right|_{E=\sqrt{p^2+m^2}},
    \\
    \chi_q
    \equiv 
    \frac{\partial^2 P}{\partial\mu_q^2},
    \quad
    \chi_B
    \equiv 
    \frac{\partial^2 P}{\partial\mu_B^2},
    \quad
    \chi_{Bq}
    \equiv 
    \frac{\partial^2 P}{\partial\mu_B\partial\mu_q},
    \label{eq : succeptibilities}
\end{gathered}
\end{equation}
where $P$ is the pressure of the SM plasma up to second order with respect to chemical potentials.
$\chi$ is the susceptibility for leptons and can be well approximated by integration of Fermi-Dirac distributions. 
On the other hand, $\chi_B$, $\chi_q$, and $\chi_{Bq}$ are related to the QCD thermodynamics and cannot be approximated by integration of the Fermi-Dirac distributions of free fermions.
Thus, we fit the numerical results obtained by Ref.~\cite{Venumadhav:2015pla}, whose fitting formulae are presented in Appendix~\ref{append : susceptibilities}.
By solving Eqs.~\eqref{eq : def of La}, \eqref{eq : nB and nQ and na}, and \eqref{eq : succeptibilities} in terms of $\mu_B$, $\mu_q$ and $\mu_a$,
we can express $V_a$ given by Eq.~\eqref{eq : finite density potential} using $L_a$.
The full expression of $V_a$ is presented in Appendix~\ref{append : relation between La and Va}.

\section{Numerical setup}
\label{sec : numerical setup}

In this section, we explain our scheme for the numerical calculation and then discuss the scan of the model parameters that can explain all DM by resonantly produced sterile neutrinos.

\subsection{Parametrization of the momentum variable}
\label{subsec : discritization method}

Before going on to the analysis, we define a dimensionless momentum mode by
\begin{equation}
    \epsilon\equiv
    \left(
    \frac{g_{*,s}(T_i)}{g_{*,s}(T)}
    \right)^{1/3}
    \frac{p}{T},
    \label{eq : def of epsilon}
\end{equation}
where $T_i$ is the initial temperature in numerical simulations, and $g_{*,s}$ is the total relativistic degrees of freedom for entropy density.%
\footnote{
    Here, we neglected the dependence of the entropy density on the finite chemical potential of the active neutrinos.
}
Due to the total entropy conservation during the radiation-dominated era, $\epsilon$ can be treated as the time-independent momentum variable. 
On the other hand, the physical dimensionless mode, defined by $\epsilon_{\rm{phys}}\equiv p/T$, depends on the temperature.

First, we consider the resonance condition. Using the above definition of $\epsilon_{\rm{phys}}$ and Eq.~\eqref{eq : finite density potential}, the resonance condition~\eqref{eq : resonance condition 1} is rewritten as
\begin{align}
    1-\frac{4\sqrt{2}\pi^2 g_{*}(T) G_{\rm{F}}}
    {45}
    \frac{\epsilon_{\rm{phys}}(T) \mathcal{L}_a T^4}
    {m_{\nu_s}^2}
    +
    2 B_a
    \frac{\epsilon_{\rm{phys}}^2(T) T^6 }{m_{\nu_s}^2}
    =
    0
    \ .
\end{align}
With $m_{\nu_s}\sim\mathcal{O}(10\,\text{--}\,100)$\,keV and $\mathcal{L}_{a}\gtrsim \mathcal{O}(10^{-5})$, which we will study below, there are two modes that satisfy the above equation for fixed $T$, which we denote as $\epsilon=\epsilon_{\rm{low}}$ and $\epsilon_{\rm{high}}$ with $0 < \epsilon_{\rm{low}}<\epsilon_{\rm{high}}$. 
But the contribution from the production at $\epsilon=\epsilon_{\rm{high}}$ is subdominant because the coherence between two states is highly suppressed due to the thermal damping effect.
The lower resonant mode, $\epsilon_{\rm{low}}$, which plays the dominant role for the resonant production, is given by
\begin{align}
    \epsilon_{\rm{low}}(T,\mathcal{L}_a(T))
    = &
    \left(
    \frac{g_{*,s}(T_i)}
    {g_{*,s}(T)}
    \right)^{1/3}
    \nonumber \\
    & \times 
    \frac{
    2\pi^2G_{\rm{F}}g_{*,s}(T)\mathcal{L}_aT^4
-\sqrt{4\pi^4G_{\rm{F}}^2g_{*,s}(T)^2\mathcal{L}_a^2T^8
-2025B_am_{\nu_s}^2T^6}
    }{45\sqrt{2}B_aT^6}
    .
    \label{eq : epsilon low}
\end{align}
The resonant $\nu_s$ production leads to the time evolution of $\mathcal{L}_a$, which in turn backreacts on the production rate of $\nu_s$. 
The time evolution of $\mathcal{L}_a$ is dominantly driven by the modes with extremely narrow momentum widths around the resonance $\epsilon\simeq \epsilon_{\rm{low}}$. 
Therefore, accurately evaluating this evolution requires fine momentum discretization around $\epsilon=\epsilon_{\rm{low}}$. 
When we focus on a fixed $\epsilon$, the resonance occurs first at $T = T_{\rm{high}}$ satisfying $\epsilon = \epsilon_\mathrm{high}$, and then at $T = T_{\rm{low}}$ satisfying $\epsilon = \epsilon_\mathrm{low}$.
The parameter dependence of $T_{\rm{low}}$, which plays a dominant role for resonant production, is approximately given by
\begin{align}
    T_{\rm{low}}
    \propto 
    \frac{m_{\nu_s}^{1/2}}{\mathcal{L}_a^{1/4}\epsilon^{1/4}}.
    \label{eq : resonance cond for T}
\end{align}
In our previous work~\cite{Kasai:2024diy}, we employed a linear discretization with $\mathcal{O}(10^4)$ bins over the range 
$\epsilon_{\rm{min}}<\epsilon<\epsilon_{\rm{max}}$, where $\epsilon_{\rm{min}}\simeq 0$ and $\epsilon_{\rm{max}}=20$. 
However, this resolution remains insufficient, especially when the initial value of $L_a$, $L_a^{\rm{init}}$, is as large as $L_a^{\rm{init}}\gtrsim 10^{-3}$.
However, a larger number of momentum bins would significantly increase the computational cost. 

Therefore, in this paper, we take a more sophisticated scheme for the binning of the momentum space.
Since $\epsilon_{\rm{low}}$ depends on time, we adopt a dynamical discretization scheme in which the momentum bins are always most finely spaced around the instantaneous resonant mode~\cite{Kainulainen:2001cb, Hannestad:2013pha}.
First, we define a variable $u$ by
\begin{align}
    u(\epsilon)\equiv \frac{\epsilon-\epsilon_{\rm{min}}}{\epsilon_{\rm{max}}-\epsilon_{\rm{min}}},
    \label{eq : def of u}
\end{align}
such that $\epsilon_{\rm{min}}<\epsilon<\epsilon_{\rm{max}}$ corresponds to $0<u<1$. 
Furthermore, to finely discretize the modes around the resonance, we introduce a new variable $v(T)$ defined by 

\begin{align}
    u(v,T)
    =
    u_{\rm{res}}(T)+
    \alpha\left(v-v_{\rm{res}}(T)\right)
    +
    \beta(T)\left(v-v_{\rm{res}}(T)\right)^3,
    \label{eq : relation between u and v}
\end{align}
where we have defined 
\begin{align}
    u_{\rm{res}}(T)
    \equiv 
    \frac{\epsilon_{\rm{low}}(T)-\epsilon_{\rm{min}}}{\epsilon_{\rm{max}}-\epsilon_{\rm{min}}}.
    \label{eq : def of u res}
\end{align}
Here, we have introduced a small parameter $\alpha>0$ that controls the momentum resolution near the resonance point $\epsilon=\epsilon_{\rm{low}}$. 
In particular, the momentum around $\epsilon_\mathrm{low}$ is discretized with intervals smaller by a factor of $\sim \alpha$ compared to linear binning with the same number of bins.
In the following, we take $\alpha=10^{-3}$, which is sufficient to achieve good numerical precision.
We choose $\beta(T)\, (>0)$ and $v_{\rm{res}}(T)$ such that the range $0<u<1$ exactly corresponds to $0<v<1$. 
Under this condition, both $v_{\rm{res}}(T)$ and $\beta(T)$ are determined analytically. 
Their concrete expressions are presented in Appendix~\ref{Append : beta and vres}.

From now on, we adopt $(v,T)$ as the basic variables of the density matrix equation instead of $(\epsilon,T)$. 
The bin variable $v$ is then linearly discretized over the range $0<v<1$ as
\begin{align}
    v=\frac{i-1}{N_{\rm{bin}}-1}
    \;\;
    (1\leq i\leq N_{\rm{bin}}),
\end{align}
where $N_{\rm{bin}}$ is the number of bins used for numerical computations. 

\subsection{Stationary point approximation}
\label{subsec : stationary point approximation}

In the numerical computation, we replace $\rho_{as}$ by the stationary solution of Eq.~\eqref{eq : second line in density operator equation} given by
\begin{align}
    \rho_{as}
    \simeq
    \rho_{as,{\rm{stat}}}
    \equiv
    -\frac{\sin2\theta}{2\cos2\theta-\frac{2p}{m_{\nu_s}^2}(2V_a-i\Gamma_{\nu_a})}
    \rho_{aa}.
    \label{eq : rho as stat}
\end{align}
We call this approximation the stationary point approximation.
Here, we used $|\rho_{aa}|\gg |\rho_{ss}|$ for the parameter range of our interest. 
By substituting Eq.~\eqref{eq : rho as stat} into Eqs.~\eqref{eq : first line in density operator equation}, and~\eqref{eq : time evolution of L}, we obtain
\begin{align}
    \frac{{\rm{d}}}{{\rm{d}}t} 
    L_a(T)
    &=
    -\frac{45}{4\pi^4g_{*,s}(T_i)}\int{{\rm{d}}\epsilon}\;\epsilon^2 
    \Gamma_{\nu_a}(p,T)
    \left[
        \theta_M^2(\epsilon,T)
        f_{\nu_a}(\epsilon,T)
        -\bar{\theta}_M^2(\epsilon,T)
        f_{\bar{\nu}_a}(\epsilon,T)
    \right]
    \ ,
    \label{eq : basic equation L_asym}
    \\
    \frac{{\rm{d}}}{{\rm{d}}t} f_{\nu_s}(\epsilon, T)
    &=
    \Gamma_{\nu_a}(p,T)
    \left(
        \theta_M^2(\epsilon,T)
        f_{\nu_a}(\epsilon,T)
        +
        \bar{\theta}_M^2(\epsilon,T)
        f_{\bar{\nu}_a}(\epsilon,T)
        \right)
        \ .
    \label{eq : basic equation f}
\end{align}
Here, we have defined $f_{\nu_a}\equiv\rho_{aa}$, $f_{\bar{\nu}_a}\equiv\bar{\rho}_{aa}$, $f_{\nu_s}\equiv\rho_{ss}+\bar{\rho}_{ss}$, and $\theta_M$ and $\bar{\theta}_M$ are effective mixing angles in the thermal plasma given by
\begin{equation}
\begin{aligned}
   \theta_M^2(\epsilon,T)
   &\equiv
    \theta^2
    \left[
        \left( 1 - \frac{2p}{m_s^2}V_{a}(\epsilon,T) \right)^2
        + \frac{p^2\Gamma_{\nu_a}^2(\epsilon,T)}{m_{\nu_s}^4}
    \right]^{-1}
    \ ,
   \\
    \bar{\theta}_M^2(\epsilon,T)
   &\equiv
    \theta^2
    \left[
        \left(1 - \frac{2p}{m_s^2}V_{\bar{a}}(\epsilon,T) \right)^2
        + \frac{p^2\Gamma_{\nu_a}^2(\epsilon,T)}{m_{\nu_s}^4}
    \right]^{-1}
    \ .
    \label{definition of thetaM}
\end{aligned}
\end{equation}
Here, we used the approximations of $\cos2\theta\simeq 1$ and $\sin2\theta\simeq 2\theta$, and used $f_{\nu_a}, f_{\bar{\nu}_a}\gg f_{\nu_s}$, which hold for the parameter region of our interest. Here, we comment on the comparisons with some existing literature. 
This formula agrees with Ref.~\cite{Ghiglieri:2015jua}, our previous work~\cite{Kasai:2024diy} and the very recent paper~\cite{Akita:2025txo}. On the other hand, there is a slight difference from Refs.~\cite{Abazajian:2001nj,Kishimoto:2008ic, Venumadhav:2015pla}, where they start from the picture of neutrino oscillations and obtain an additional term $\propto $ $\sin^22\theta$ in the denominator in Eq.~\eqref{definition of thetaM}.
This difference becomes relevant in the case with $L_a^{\rm{init}}\gtrsim \mathcal{O}(10^{-3})$, where the value of the damping factor ${p^2\Gamma_{\nu_a}^2(\epsilon,T)}/{m_{\nu_s}^4}$ at the resonance becomes comparable to $\sin^22\theta$, as discussed in detail in Ref.~\cite{Akita:2025txo}.

Here, let us discuss the validity of the stationary point approximation.
The exact solution of Eq.~\eqref{eq : second line in density operator equation} is given by the damped oscillations around $\rho_{as,{\rm{stat}}}$ with damping timescale of $\Gamma_{\nu_a}^{-1}$.
Away from the resonance, this provides a very good approximation, since the damping timescale $\Gamma_{\nu_a}^{-1}$ is much shorter than the timescale for the evolution of $\rho_{as,{\rm{stat}}}$, which is $\mathcal{O}(H^{-1})$. 
On the other hand, during the resonance, $\rho_{as,{\rm{stat}}}$ undergoes a rapid time evolution. 
The above approximation is valid if the resonance width $\delta t_{\rm{res}}$ for a given mode, which can be estimated by~\cite{Kasai:2024diy}
\begin{equation}
    \delta t_{\rm{res}}
    \simeq
    \left.
        \frac{{\rm{d}}t}{{\rm{d}}T}
    \right|_{T_{\rm{res}}}
    \times 
    \frac{\Gamma_{\nu_a}(T_{\rm{res}})T_{\rm{res}}}{3V_a(\epsilon_{\rm{phys}},T_{\rm{res}})}
    \ ,
    \label{eq : during resonance condition3}
\end{equation}
satisfies $\delta t_{\rm{res}}\gg \Gamma_{\nu_a}^{-1}$.
Here, we define a parameter $D_{\rm{stat}}$ by
\begin{equation}
    D_{\rm{stat}}
    \equiv
    \left.
        \frac{{\rm{d}}t}{{\rm{d}}T}
    \right|_{T_{\rm{low}}}
    \times
    \frac{\Gamma_{\nu_a}^2(T_{\rm{low}})T_{\rm{low}}}
    {3V_a(\epsilon_{\rm{phys}}, T_{\rm{low}})}
    \ .
    \label{eq : during resonance condition4}
\end{equation}
Then, $D_{\rm{stat}}\gg1$ is the sufficient condition for the stationary point approximation to be accurate.

In our previous paper~\cite{Kasai:2024diy}, we have expected that the above approximation would be violated with smaller $m_{\nu_s}$ and larger $L_a^{\rm{init}}$, with which $D_{\rm{stat}}$ becomes smaller than unity.
However, we have numerically checked that this approximation hardly changes the final abundance of the sterile neutrino in the parameter region where $D_\mathrm{stat} < 1$ by directly solving the kinetic equation, which is described in Appendix~\ref{append : full QKE}.%
\footnote{
    Ref.~\cite{Akita:2025txo} also numerically examines in detail the parameter region with $D_{\rm{stat}}< 1$ while neglecting the time evolution of $L_a$ and find that the stationary point approximation still works even for $D_{\rm{stat}}< 1$.  
    Moreover, Ref.~\cite{Akita:2025txo} shows that the Boltzmann equation~\eqref{eq : basic equation f} is more accurate than that with $\sin^2 2\theta$ term.
}
Thus, we adopt Eqs.~\eqref{eq : basic equation f} and~\eqref{definition of thetaM} in the parameter space of our interest.

Now, we dynamically discretize $\epsilon$ as explained in Sec.~\ref{subsec : discritization method}. 
By using $v$, we can rewrite the basic equation~\eqref{eq : basic equation L_asym} as
\begin{align}
    \frac{{\rm{d}}}{{\rm{d}}t} 
    L_a(T)
    =
    -\frac{45}{4\pi^4g_{*,s}(T_i)}\int{\rm{d}}v
    \left.
    \frac{\partial\epsilon}{\partial v}
    \right|_{T}
    \epsilon^2 
    \Gamma_{\nu_a}(\epsilon,T)
    \left[
        \theta_M^2(\epsilon,T)
        f_{\nu_a}(\epsilon,T)
        -
       \bar{\theta}_M^2(\epsilon,T)
        f_{\bar{\nu}_a}(\epsilon,T)
    \right]
    \ ,
    \label{eq : mode equations}
\end{align}
where $\epsilon$ is now a function of $v$ and $T$ explicitly given by
\begin{align}
    \epsilon(v,T)\equiv 
    (\epsilon_{\rm{max}}-\epsilon_{\rm{min}})u(v,T)
    +\epsilon_{\rm{min}}.
\end{align}
Notice that, since we have approximated $f_{\nu_a}\gg f_{\nu_s}$, the time derivative of $L_a$ (Eq.~\eqref{eq : mode equations}) is determined by the cosmic temperature and $\mathcal{L}_a$, independently of $f_s(\epsilon,T)$.
Thus, Eq.~\eqref{eq : mode equations} is a closed equation. 
Therefore, we first numerically solve Eq.~\eqref{eq : mode equations} to determine the time evolution of $L_a$, and then substitute it to Eq.~\eqref{eq : basic equation f} to numerically solve the evolution of each momentum mode of $\nu_s$.
We have checked the consistency between the time evolution of $L_a$ obtained in the first step and that calculated from $f_{\nu_s}(\epsilon,T)$ obtained in the second step.

\subsection{Results}
\label{subsec : Results with stationary approx}

\begin{figure}[t]
\centering
\begin{tabular}{c} 
\includegraphics[width=0.7\linewidth]{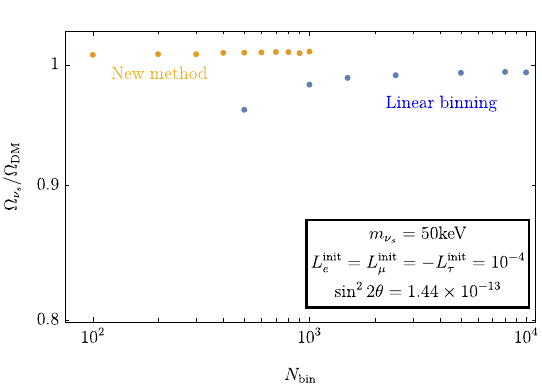}
\\
\includegraphics[width=0.7\linewidth]{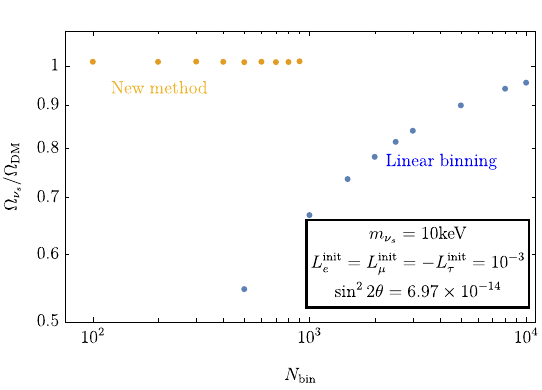}
\end{tabular}
    \caption{Dependence of the final abundance of sterile neutrinos on $N_{\rm{bin}}$. 
    The orange dots show the results of the method explained in Sec.~\ref{subsec : discritization method}.
    Here, we choose $\theta$ so that the sterile neutrinos account for all DM with $N_{\rm{bin}}=500$.
    For comparison, we show the results of the previous method by the blue dots, which imply that even $N_{\rm{bin}}\sim 10^4$ is insufficient for the accurate numerical computation in the previous method.
    On the other hand, the orange dots imply that $N_{\rm{bin}} \gtrsim 100$ is sufficient in the current method.
    }
    \label{fig : comparison with previous method}
\end{figure}

In this subsection, we focus on the case with $L_e^{\rm{init}}=L_\mu^{\rm{init}}=-L_\tau^{\rm{init}}$, which can be realized within Affleck-Dine leptogenesis, as discussed in Ref.~\cite{Kasai:2024diy}. 
Other cases can be discussed in the same way, and we summarize the results in another intriguing example in Sec.~\ref{subsec : the case with vanishing lepton charge in total}. 
We also set the mixing between electron neutrino and sterile neutrino, i.e., $a=e$. 
The results would not be changed significantly even when we choose another flavor, as long as the resonance plays a dominant role, as discussed in Ref.~\cite{Kasai:2024diy}.

First, as the validation of our numerical scheme, we show the dependence of the final $\nu_s$ abundance on $N_{\rm{bin}}$ for $(m_{\nu_s}, L_{e}^{\rm{init}}) = (50\,\mathrm{keV},\, 10^{-4})$ and $(10\,\mathrm{keV},\, 10^{-3})$ in Fig.~\ref{fig : comparison with previous method}. 
In the figure, the orange dots show the results with the method presented in Sec.~\ref{subsec : discritization method}, and the blue dots represent the results using linear binning over $0<\epsilon<20$, as implemented in Ref.~\cite{Kasai:2024diy}. 
This figure implies that the method in Sec.~\ref{subsec : discritization method} achieves numerical convergence at $N_{\rm{bin}} \gtrsim 100$, whereas the previous method is insufficient even with $N_{\rm{bin}}=10^4$. 

The linear binning approach deteriorates for smaller $m_{\nu_s}$ and larger $L_a^{\rm{init}}$. 
For smaller $m_{\nu_s}$, this is because a larger number of $\nu_s$ must be produced to reach the DM abundance, leading to a larger change in $\mathcal{L}_a$. 
Consequently, the evolution of $L_a$ becomes more significant with smaller $m_{\nu_s}$.
For larger values of $L_a^{\rm{init}}$, the resonance becomes narrower, and consequently, only a smaller range of momentum modes contributes to the evolution of lepton asymmetry. Indeed, too large $\Delta\epsilon$ underestimates the final sterile neutrino abundance.
\begin{figure}[t]
\centering
\begin{tabular}{c} \includegraphics[width=0.7\linewidth]{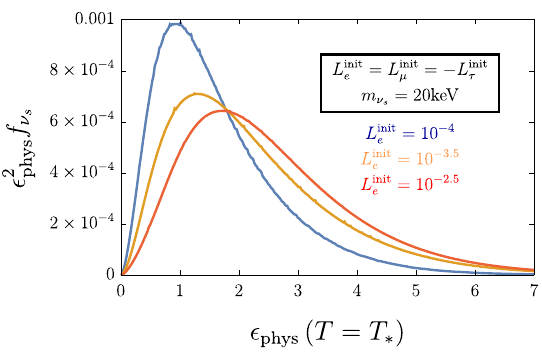}
\\
\includegraphics[width=0.7\linewidth]{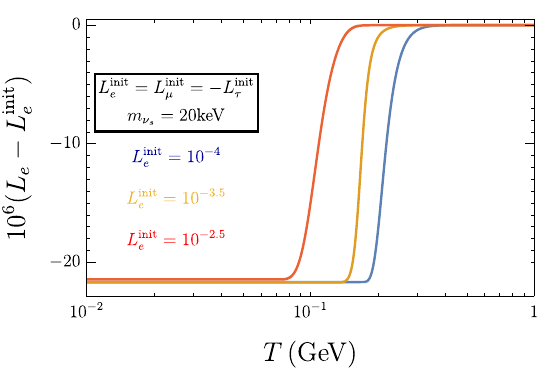}
\end{tabular}
\caption{Upper panel: Momentum distributions of produced sterile neutrinos with $\epsilon_{\rm{phys}}$ evaluated at $T=T_* \equiv 1$\,MeV for $L_e^{\rm{init}}=L_{\mu}^{\rm{init}}=-L_\tau^{\rm{init}}$. 
The blue, orange, and red lines correspond to $L_e^{\rm{init}}=10^{-4},10^{-3.5},10^{-2.5}$, respectively. 
For a larger value of $L_{e}^{\rm{init}}$, the peak of the spectrum lies in a higher momentum mode. 
Lower panel: Time evolution of lepton asymmetry with the same parameters as in the upper panel.
}
\label{fig : plot of spectra 1}
\end{figure}

In Fig.~\ref{fig : plot of spectra 1}, we show the result of the final momentum spectra of the sterile neutrinos for $m_{\nu_s}=20$\,keV and several values of $L_e^{\rm{init}}$, where $\epsilon_{\rm{phys}}$ is evaluated at $T = T_* \equiv 1\,{\rm{MeV}}$. 
Here, we use $N_{\rm bin}=500$ and choose $\theta$ so that the sterile neutrinos account for all DM. 
It is seen that the peak momentum of the final spectrum increases as $L_{e}^{\rm{init}}$, which is almost the same as the results presented in Ref.~\cite{Kasai:2024diy}. 

\begin{figure}[t]
    \centering
    \begin{tabular}{c} 
    \includegraphics[width=0.85\linewidth]{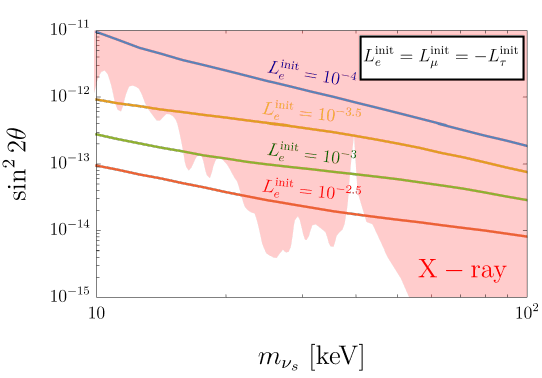}
    \end{tabular}
    \caption{Contours of $m_{\nu_s}$ and $\sin^2 2\theta$ to explain all DM for $L_e^\mathrm{init} =10^{-4}$ (blue), $10^{-3.5}$ (orange), $10^{-3}$ (green), and $10^{-2.5}$ (red) for $L_e^{\rm{init}}=L_{\mu}^{\rm{init}}=-L_\tau^{\rm{init}}$.
    The light-red shaded region shows the X-ray constraint, where we use the $2\sigma$ constraint from \textit{NuSTAR} project for $m_{\nu_s}<40\,$keV~\cite{Krivonos:2024yvm} and $2\sigma$ constraint from INTEGRAL project for $m_{\nu_s}>40$\,keV~\cite{Fischer:2022pse}.
   }
    \label{fig : contour without Lyman alpha}
\end{figure}

Furthermore, in Fig.~\ref{fig : contour without Lyman alpha}, we show the relation between $m_{\nu_s}$ and $\theta$ to realize all DM for several values of $L_{e}^{\rm{init}}$ adopting $N_{\rm{bin}}=500$. 
Here, for a given $m_{\nu_s}$, we perform a numerical search scanning over $\theta$ and find the value of $\theta$ with which the final abundance satisfies $|(\Omega_{\nu_s}-\Omega_{\rm{DM}})/\Omega_{\rm{DM}}|<0.01$ ($\Omega_{\nu_s\,(\mathrm{DM})}$: the density parameter of $\nu_s$ (DM)). 
There are some minor differences between the contours in Fig.~\ref{fig : contour without Lyman alpha} and those presented in Ref.~\cite{Kasai:2024diy}, in which a linear discretization scheme with $N_{\rm{bin}}=10^4$ was adopted and the redistribution of asymmetry to the charged leptons is neglected, although the overall behavior is the same. 

\subsection{Free-streaming length of the sterile neutrinos}

Here, we discuss the free streaming of the produced sterile neutrinos. 
Since the produced sterile neutrinos can have non-negligible momenta, the fluctuations of the matter density are suppressed on small scales.
From the observations of the Lyman-$\alpha$ forests in the quasar spectra, the amplitude of the matter power spectrum on small scales can be constrained. 
There also exists another constraint from the structure formation by observations of strong lensing and galaxy distributions~\cite{Zelko:2022tgf}.
In this paper, for simplicity, we assume that the suppression of small-scale density perturbations is represented by the free-streaming length of the sterile neutrinos.
Thus, we calculate the free-streaming lengths from the final spectrum of the sterile neutrinos and compare them with the observational constraints.
The constraint on the free-streaming length of DM, denoted as $\lambda_{\rm{FS}}$, can be inferred by using the constraint on the mass of early-decoupled thermal relic~\cite{Viel:2005qj, Baur:2015jsy, Baur:2017stq} and is given by $\lambda_{\rm{FS}}^{\rm{upper}}\lesssim 0.129$\,Mpc for the Lyman $\alpha$ and $\lambda_{\rm{FS}}^{\rm{upper}}\lesssim 0.034$\,Mpc for the galaxy counting $+$ strong lensing, based on the method discussed in Ref.~\cite{Kasai:2024diy}.

The typical free-streaming length of particles decoupled from the thermal plasma is given by~\cite{Baur:2017stq}
\begin{align}
   \lambda_{\rm{FS}}
   \simeq 
   a(t)\int_{a_*}^{a(t)}{\rm{d}}a\frac{v(\langle\epsilon\rangle,t)}{a^2(t)H(t)}
   \ ,
   \label{eq : free streaming length}
\end{align} 
where $\langle \epsilon \rangle$ is $\epsilon$ averaged over the momentum distribution at $T = T_* = 1$\,MeV, which is lower than both of the decoupling temperature and the particle production temperature, and $a_*$ is the scale factor at $T = T_*$.
The velocity $v(t)$ is given by 
\begin{align}
    v  
    &=
    \frac{p}{\sqrt{p^2+m_{\nu_s}^2}}
    =
    \frac{\epsilon_{\rm{phys}}(T_{*})}
    {\sqrt{ \epsilon_{\rm{phys}}(T_{*})^2+(\frac{m_{\nu_s}}{T_{*}})^2(\frac{a}{a_*})^2 }}
    \ ,
    \label{eq : velocity and momentum}
\end{align}
where we used $p = \epsilon_{\rm{phys}}(T_{*}) T_{*} a_*/a$. 
Here, we neglected the free-streaming length before $T=T_*$. 
\begin{figure}[t]
    \centering
    \begin{tabular}{c} 
    \includegraphics[width=1.0\linewidth]{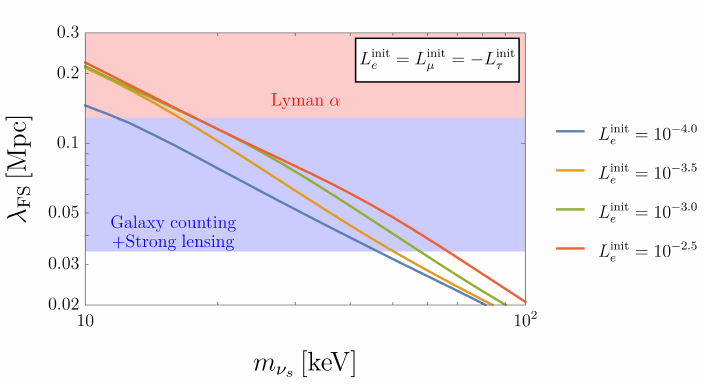}
    \end{tabular}
    \caption{Dependence of the free-streaming length of sterile neutrinos $\lambda_{\rm{FS}}$ on $m_{\nu_s}$ for $L_e^{\rm{init}}=L_{\mu}^{\rm{init}}=-L_\tau^{\rm{init}}$. 
    Different colors correspond to different initial values of the lepton asymmetry $L_e^{\rm{init}}$. 
    The vacuum mixing angle $\theta$ is fixed to match the DM abundance in each $L_e^{\rm{init}}$ and $m_{\nu_s}$. 
    The red region is disfavored by the Lyman $\alpha$ constraint $m_{\rm{th}}\gtrsim 3.3$\,keV~\cite{Zelko:2022tgf}. The violet region is disfavored by more stringent constraint from the combinations of strong lensing and galaxy counting obtained by Ref.~\cite{Zelko:2022tgf}, which argues $m_{\rm{th}}\gtrsim 9.8$\,keV.
    }
    \label{fig : free streaming length with Le = Lmu = -Ltau}
\end{figure}

\begin{figure}[t]
\centering
\begin{tabular}{c} 
\includegraphics[width=0.8\linewidth]{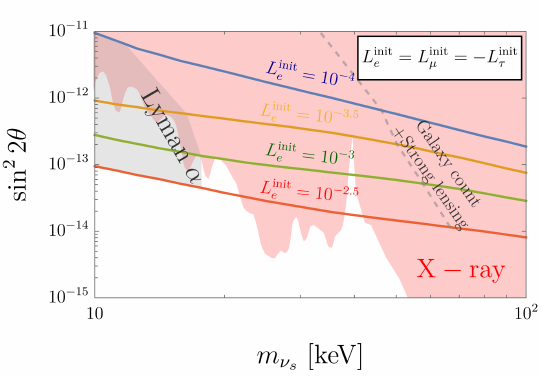}
\end{tabular}
    \caption{$m_{\nu_s}$ and $\sin^2 2\theta$ to explain all DM with the observational constraints for  $L_e^{\rm{init}}=L_{\mu}^{\rm{init}}= -L_\tau^{\rm{init}}$. 
    The light shaded region is excluded by X-ray observations. 
    The gray shaded region is disfavored by the Lyman-$\alpha$ observations~\cite{Zelko:2022tgf}, and the gray dashed line is the 2$\sigma$ boundary of the combination of constraints from galaxy counting and strong lensing~\cite{Zelko:2022tgf}. 
    }
    \label{fig : contour including Lyman alpha}
\end{figure}

In Fig.~\ref{fig : free streaming length with Le = Lmu = -Ltau}, we show the free-streaming length of the sterile neutrinos as a function of $m_{\nu_s}$ for a given $L_e^{\rm{init}}$, using the momentum distribution obtained in Sec.~\ref{subsec : Results with stationary approx}. 
By requiring $\lambda_{\rm{FS}}(m_{\nu_s}) < \lambda_{\rm{FS}}^{\rm{upper}}$, we obtain the constraint on $m_{\nu_s}$ for a given $L_a^{\rm{init}}$. 
In Fig.~\ref{fig : contour including Lyman alpha}, we show the contour plot to explain all DM for $L_e^{\rm{init}}=L_{\mu}^{\rm{init}}=-L_\tau^{\rm{init}}$ together with constraints from the structure formation and X-ray observations. 

We conclude that the lepton-to-entropy ratio must be $\gtrsim 10^{-3}$ to evade the X-ray constraints. 
Furthermore, the Lyman-$\alpha$ constraints on thermal relic DM~\cite{Zelko:2022tgf} suggest that the mass of sterile neutrinos must be $m_{\nu_s}\gtrsim 20$\,keV. If we use the most stringent constraint from structure formation by a combination of strong lensing and galaxy counting~\cite{Zelko:2022tgf}, the constraint becomes further stringent. 
However, since we just compared the averaged free-streaming lengths, this constraint would be altered if we studied in detail the effect on the matter power spectrum using our sterile neutrino spectra as done in Refs.~\cite{Baur:2015jsy, Vogel:2025aut}.
We also note that these constraints will be relaxed if a fraction of sterile neutrinos is produced during the injection process of lepton number density, as we will discuss in Sec.~\ref{sec : generation of a large lepton asymmetry}.

\subsection{Case with vanishing total lepton asymmetry}
\label{subsec : the case with vanishing lepton charge in total}

As another interesting possibility, we investigate the sterile neutrino production when the lepton asymmetry of each flavor is large but the total asymmetry vanishes, which is suggested by the recent papers~\cite{Vogel:2025aut,Akita:2025txo}.
In this case, the flavor oscillations can reduce the asymmetry in each flavor after the sterile neutrino production and before the onset of BBN, which significantly relaxes the constraints on the lepton asymmetry from BBN. 
Such a lepton flavor asymmetry can be realized within Affleck-Dine leptogenesis~\cite{Akita:2025zvq}, when we consider the dynamics of the flat directions in the minimal supersymmetric standard model, such as $QuLe$. 

\begin{figure}[t]
\centering
\begin{tabular}{c} 
\includegraphics[width=0.8\linewidth]{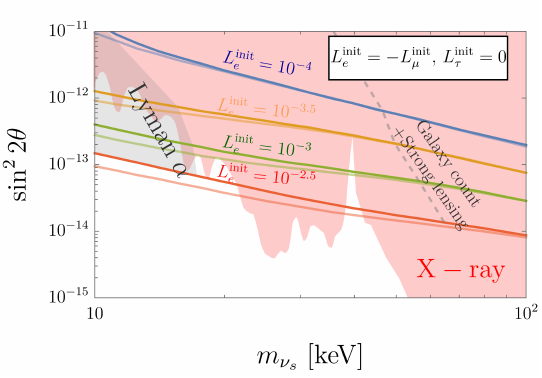}
\end{tabular}
    \caption{Same as Fig.~\ref{fig : contour including Lyman alpha} except for $L_e^{\rm{init}}=-L_\mu^{\rm{init}}$ and $L_\tau^{\rm{init}}=0$.
    For comparison, we also show the contours in Fig.~\ref{fig : contour including Lyman alpha} by the lighter colors.
    }
    \label{fig : contour including Lyman alpha with net zero lepton charge}
\end{figure}

In this paper, as an example, we focus on the case where $L_e^{\rm{init}}=-L_{\mu}^{\rm{init}}$ and $L_{\tau}^{\rm{init}}=0$. 
The computational method is the same as in the previous case, and we show the $m_{\nu_s}$--\ $\theta$ contour plot to explain all DM in Fig.~\ref{fig : contour including Lyman alpha with net zero lepton charge}. 
For comparison, we also show the results in the previous case with the lighter colors.
We see that the required mixing angle to realize the observed DM abundance in this case is larger than in the previous setup. 
This is because the value of $\mathcal{L}_e$ is smaller in this case for the same value of $L_e^{\rm init}$. 
Since the evolution of $\mathcal{L}_e$ depends on the redistribution of asymmetry, the difference in the required mixing angle nontrivially depends on $m_{\nu_s}$ and $\sin^2 2\theta$.

\section{Generation of a large lepton asymmetry}
\label{sec : generation of a large lepton asymmetry}

In this section, we discuss the generation of lepton asymmetry within the framework of Affleck-Dine (AD) leptogenesis.

\subsection{Decay of Q-balls}

As a Q-ball formation model, we consider the gauge-mediated SUSY breaking scenario.
In AD leptogenesis, lepton asymmetry is generated from the coherent phase rotation of a complex field $\phi$, called an AD field, carrying lepton number. 
The resulting lepton number density
is written as
\begin{equation}
    n_L(t_\mathrm{osc}) \simeq \delta_\mathrm{eff} m_\phi \varphi_\mathrm{osc}^2 
    \ ,
    \label{resulting lepton number in AD leptogenesis}
\end{equation}
where $m_\phi$ is the mass of sleptons, and $\delta_\mathrm{eff}$ is the efficiency parameter for asymmetry production.

After the AD field starts to oscillate at $t=t_\mathrm{osc}$, the potential for the AD field is given by
\begin{align}
    V
    &=
    V_{\rm{gauge}} + V_{\rm{gravity}}
    \ .
    \label{gauge mediation and gravity mediation potential}
\end{align}
Here, $V_{\rm{gauge}}$ is the contribution from gauge-mediation given by~\cite{deGouvea:1997afu,Kusenko:1997si}
\begin{align}
    V_{\rm{gauge}}
    =
    M_{\rm{F}}^4 \left( \ln \frac{|\phi|^2}{M_S^2} \right)^2
    \ ,
    \label{eq:Vgauge}
\end{align}
where $M_{\rm{F}}$ is the SUSY breaking scale, and $M_S$ is the scale of the messenger sector.
Here, we assumed $\varphi \equiv |\phi|\gg M_S$.
$V_{\rm{gravity}}$ is the contribution from gravity-mediation given by~\cite{deGouvea:1997afu,Kusenko:1997si}
\begin{align}
    V_{\rm{gravity}}
    =
    m_{3/2}^2
    \left[
        1 + K\ln\left( \frac{|\phi|^2}{M_{\rm{Pl}}^2} \right)
    \right]
    |\phi|^2
    \ ,
    \label{eq:Vgravity}
\end{align}
where $m_{3/2}$ is the mass of gravitino, $M_\mathrm{Pl}$ is the reduced Planck mass, $K$ is a dimensionless constant, which we assume to be positive in this paper.
We also assume that $V_{\rm{gravity}}$ is dominant in the potential when the AD field starts to oscillate, i.e., $\varphi_\mathrm{osc}  \equiv \varphi(t_\mathrm{osc}) > \varphi_{\rm{eq}}\equiv M_{\rm{F}}^2/m_{3/2}$.
When the potential is dominated by $V_{\textrm{gravity}}$ with $K>0$, the coherent oscillation of the AD field is stable against spatial fluctuations.
However, right after $\varphi$ becomes comparable to $\varphi_{\rm{eq}}$, $V_{\rm{gauge}}$ controls the dynamics of the AD field.
In this period, the AD field experiences spatial instabilities and forms spherically symmetric non-topological solitons called Q-balls~\cite{Enqvist:1997si,Kasuya:1999wu,Kasuya:2014ofa}.
We call this type of Q-ball formation scenario ``delayed-type'' Q-ball scenario~\cite{Kasuya:2001hg}.

Delayed-type Q-balls have the same properties as the gauge-mediated type Q-balls.
Then, the initial charge, mass, radius, and energy per charge of a delayed-type Q-ball are given by~\cite{Kasuya:2014ofa}
\begin{align}
    |Q_{\rm{G}}^{\mathrm{init}}|
    & \simeq
    \beta_{\rm{G}} \frac{\varphi_{\rm{eq}}^4}{M_{\rm{F}}^4}
    \simeq
    \beta_{\rm{G}} 
    \left(
        \frac{M_{\rm{F}}}{m_{3/2}}
    \right)^4
    \ , 
    \label{Q_G}
    \\[0.4em]
    M_{\rm{Q}} & \simeq \frac{4\sqrt{2}\pi}{3}\zeta M_{\rm{F}} 
    |Q_{\rm{G}}|^{\frac{3}{4}}
    \ ,
    \label{eq : Q-ball property1}
    \\[0.4em]
    R_{\rm{Q}} & \simeq \frac{1}{\sqrt{2}\zeta}M_{\rm{F}}^{-1}
    |Q_{\rm{G}}|^{\frac{1}{4}}
    \ ,
    \label{eq : Q-ball property2}
    \\[0.4em]
    \omega_{\rm{Q}} & \simeq \sqrt{2}\pi \zeta M_{\rm{F}}
    |Q_{\rm{G}}|^{-\frac{1}{4}}
    \ ,
    \label{eq : Q-ball property3}
\end{align}
where $\beta_{\rm{G}}\simeq 6\times10^{-5}$~\cite{Kasuya:2012mh} in the case with $|\delta_\mathrm{eff}| \lesssim 10^{-1}$, which we consider in this paper,
and $\zeta$ is a numerical factor  $\simeq 2^{1/4}(c_0/\pi)^{1/2}$ where $c_0\equiv 4.8\ln (m_\phi/\sqrt{2}\omega_{\rm{Q}})+7.4$~\cite{Hisano:2001dr}.
In the following, we adopt $\zeta=3.4$, which is obtained by substituting $m_{3/2}=1$\,GeV, $M_{\rm{F}}=10^6$\,GeV and $m_\phi=10^4$\,GeV. 
The Q-balls decay into neutrinos with the decay rate $\Gamma_{\rm{Q}}$ given by~\cite{Cohen:1986ct,Kawasaki:2012gk}
\begin{equation}
    \Gamma_{\rm{Q}}
    \equiv 
    -\frac{1}{Q_\mathrm{G}} \frac{\mathrm{d}Q_\mathrm{G}}{\mathrm{d} t}
    \simeq 
    \frac{N_l}{|Q_\mathrm{G}|}\frac{\omega_{\rm{Q}}^3}{12\pi^2}4\pi R_{\rm{Q}}^2
    \ ,
    \label{Qball decay rate}
\end{equation}
where $N_l=3$ is the number of active neutrino species.

To realize the generation of a large lepton asymmetry from Q-ball decay, we assume that Q-balls dominate the energy density of the universe at Q-ball decay. 
Then, the resulting lepton-to-entropy ratio from Q-ball decay is evaluated as~\cite{Kasai:2024diy}
\begin{align}
    L_{a}^{\rm{dec}} &\simeq \frac{n_L(t_{\rm{osc}})}{m_{3/2}^2\varphi_{\rm{osc}}^2} 
    \frac{\rho_{\rm{Q}}(t_{\rm{dec}})}{s(t_{\rm{dec}})} 
    \nonumber\\
    &\simeq
    \delta_\mathrm{eff}\frac{3T_{\rm{dec}}}{4m_{3/2}}
    \ ,
    \label{Relation between etaL gravitino mass and Tdecay}
\end{align}
Here, $L_a^{\rm{dec}}$ is the total lepton asymmetry injected from the Q-ball decay, including the amount that is eventually consumed by the resonant $\nu_s$ production. If the asymmetry conversion into sterile neutrinos is negligible, it is the same as $L_a^{\rm{init}}$, as defined in Sec.~\ref{sec : resonant production}.
$\rho_{\rm{Q}}$ is the energy density of the Q-balls and their decay products, and $T_\mathrm{dec}\,(t_\mathrm{dec})$ is the cosmic temperature (time) just after Q-ball decay.
Here, we have approximated that Q-balls instantaneously decay after domination when the decay rate $\Gamma_{\rm{Q}}$ becomes equal to the Hubble parameter.
Then $T_\mathrm{dec}$ is evaluated as 
\begin{align}
    T_{\rm{dec}} 
    &=
    \left(\frac{90}{\pi^2 g_{\ast}(T_{\rm{dec}})}\right)^{1/4}\sqrt{M_{\rm{Pl}}\Gamma_{\rm{Q}}}\nonumber
    \\
    &\simeq
    2.7\,{\rm{GeV}}\times \left(\frac{g_{\ast}}{80}\right)^{-1/4}
    \left(\frac{m_{3/2}}{1\,{\rm{GeV}}}\right)^{5/2}\left(\frac{M_{\rm{F}}}{10^6\,{\rm{GeV}}}\right)^{-2}
    \ ,
    \label{Tdecay}
\end{align}
where we evaluated $\Gamma_Q$ using the initial charge $Q_\mathrm{G}^\mathrm{init}$.

\subsection{Resonant production during the decay of Q-balls}
\label{Subsec : resonance during lepton number generation}
In the previous paper~\cite{Kasai:2024diy}, we assumed that the resonant production should occur after the decay of Q-balls was completed.  
This assumption is translated into the constraint on the $M_{\rm{F}}$, which is given by
\begin{equation}
    T_{\rm{low}}
    \left(
        m_{\nu_s},L_{a}^{\rm{init}} = \delta_\mathrm{eff}\frac{3T_{\rm{dec}}}{4m_{3/2}}
    \right)
    <
    T_{\rm{dec}}
    \ .
    \label{compatible condition for resonance}
\end{equation}
Using Eqs.~\eqref{Tdecay} and~\eqref{compatible condition for resonance}, we obtain the condition for $m_{3/2}$, $M_{\rm{F}}$, and $m_s$ as
\begin{align}
    M_{\rm{F}}  \lesssim & 
    ~ 3.1\times10^6\,{\rm{GeV}}
    \left(
    \frac{g_{*}(T_{\rm{decay}})}{80}
    \right)^{-1/8}
    \left(
    \frac{g_{*}(T_{\rm{low}})}{60}
    \right)^{1/10}
    \left(
    \frac{|\delta_\mathrm{eff}|}{10^{-3}}
    \right)^{1/10} \nonumber \\
    & ~ \times \left(
    \frac{m_{3/2}}{1\,{\rm{GeV}}}
    \right)^{23/20}
    \left(
    \frac{m_{\nu_s}}{50\,{\rm{keV}}}
    \right)^{-1/5}.
    \label{compatible condition for resonance in terms of MF}    
\end{align}
However, if $M_{\rm{F}}$ is slightly larger than this value, resonant production could occur during the decay process of Q-balls. 
Here, we consider the observational constraint on such a case.

Before and during the Q-ball decay, the thermal history of the universe differs from that in the radiation-dominated era, which was assumed in the previous section.
In the following, we describe the relations used in numerical calculations for $T > T_\mathrm{dec}$.
The time dependences of the mass and the charge of the Q-balls are given by
\begin{align}
    M_{\rm{Q}}(t) 
    &= 
    M_{{\rm{Q}}}^{\rm{init}}\left(
    1-\frac{t}{t_{\rm{dec}}}
    \right)^{3/5},
    \nonumber \\
    Q(t) 
    &= 
    Q_{\rm{}G}^{\rm{init}}\left(
    1-\frac{t}{t_{\rm{dec}}}
    \right)^{4/5},
\end{align}
where $t_{\rm{dec}}$ is the physical time at which the decay process is completed, and $Q_{\rm{G}}^{\rm{init}}$, $M_{\rm{Q}}^{\rm{init}}$ are the charge and mass of the Q-balls at formation, which are given by Eqs.~\eqref{Q_G} and~\eqref{eq : Q-ball property1}.
The relation between the conformal time $\eta$ and the background densities $\rho_r$ and $\rho_m$ are given by
\begin{equation}
\begin{aligned}
    \rho_m 
    &= 
    \rho_{\rm{eq},1}
    \left(
    \frac{a_{\rm{eq},1}}{a}
    \right)^3
    \frac{M_{\rm{Q}}}{M_{\rm{Q}}^{{\rm{init}}}},
    \\
    \frac{{\rm{d}}\rho_r}{{\rm{d}}\eta}
    &=
    -4\mathcal{H}\rho_r
    -\frac{{\rm{d}}\ln M_{{\rm{Q}}}}{{\rm{d}}\eta}\rho_m.
    \label{eq : background energy density}
\end{aligned}
\end{equation}
Here, the subscript ``${{\rm{eq}},1}$'' denotes the early matter-radiation equality. 
We have approximated $Q=Q_{\rm{init}}$ at $\eta=\eta_{\rm{eq},1}$, which is valid in the parameter sets of our interest.
Furthermore, until the Q-ball decay, the relations among the scale factor $a$, conformal Hubble parameter $\mathcal{H}\equiv ({\rm{d}}a/{\rm{d}}\eta)/a$, and $\eta$ are given by
\begin{equation} 
\begin{aligned}
    \frac{a(\eta)}{a(\eta_{\rm{eq},1})}
    &=
    \left(
    \frac{\eta}{\eta_*}
    \right)^2
    +2\left(
    \frac{\eta}{\eta_*}
    \right),
    \\
    \mathcal{H}(\eta)
    &=
    \frac{2\eta+2\eta_*}{\eta^2+2\eta\eta_*},
    \label{eq : scale factor and Hubble}
\end{aligned}
\end{equation}
where $\eta_*\equiv \eta_{\rm{eq},1}/(\sqrt{2}-1)$.
Assuming that the radiation from the Q-balls is always instantaneously thermalized, the relation between the physical temperature and the radiation energy density is given by
\begin{align}
    \rho_{r}=\frac{\pi^2}{30}
    g_{*}(T)T^4,
    \label{eq : rho r}
\end{align}
where $g_{*}$ is the relativistic degrees of freedom for the energy density.
By combining Eqs.~\eqref{eq : background energy density} and \eqref{eq : rho r}, we obtain the relation between $\eta$ and $T$ as
\begin{align}
    \frac{{\rm{d}}T}{{\rm{d}}\eta}
    &=
    -
    \left(
    \frac{4}{T}
    +\frac{{\rm{d}}\ln g_{*}}{{\rm{d}}T}
    \right)^{-1}
    \left(
    4\mathcal{H}+
    \frac{{\rm{d}}\ln M_{\rm{Q}}}{{\rm{d}}\eta}
    \frac{g_{*}(T_{\rm{eq,1}})}{g_{*}(T)}
    \left(
    \frac{T_{\rm{eq,1}}}{T}
    \right)^4
    \left(
    \frac{a_{\rm{eq,1}}}{a}
    \right)^3
    \frac{M_{\rm{Q}}}{M_{\rm{Q},init}}
    \right).
    \label{eq : temperature eta relation}
\end{align}
We solve this equation to obtain $T(\eta)$. 

In this case, the comoving entropy density is not conserved due to the Q-ball decay.
Thus, we modify the definition of the dimensionless momentum mode $\epsilon$ as
\begin{align}
    \epsilon
    \equiv 
        \frac{a}{a_{\rm{dec}}}
        \frac{p}{T_{\rm{dec}}}
    \ .
\end{align}
Furthermore, we modify the definition of $L_a$ as
\begin{align}
L_a
\equiv 
\frac{n_a-n_{\bar{a}}
+
n_{\nu_a}-n_{\bar{\nu}_a}}{s_{\rm{tot}}(T_{\rm{dec}})
\left(\frac{a_{\rm{dec}}}{a}
\right)^3}.
\label{eq : def of La during injection}
\end{align}
Then, we obtain the following equations as
\begin{align}
    \frac{{\rm{d}}}{{\rm{d}}t} L_a(T)
    =&
-L_{a}^{\rm{dec}}\frac{\dot{Q}}{Q_{\rm{init}}}
\nonumber \\
    &-\frac{45}{4\pi^4g_{*,s}(T_{\rm{dec}})}\int{{\rm{d}}\epsilon}\;\epsilon^2 
    \Gamma_{\nu_a}(p,T)
    \left[
        \theta_M^2(\epsilon,T)
        f_{\nu_a}(\epsilon,T)
        \right.
        -
        \left.
\bar{\theta}_M^2(\epsilon,T)
        f_{\bar{\nu}_a}(\epsilon,T)
    \right]
    \ ,
    \nonumber \\
    \frac{{\rm{d}}}{{\rm{d}}t} 
    f_{\nu_s}(\epsilon, T)
    =&
    \Gamma_{\nu_a}(p,T)
        \left(
        \theta_M^2(\epsilon,T)
        f_{\nu_a}(\epsilon,T)
        +
        \bar{\theta}_M^2(\epsilon,T)
        f_{\bar{\nu}_a}(\epsilon,T)
        \right).
    \label{eq : basic eq during Qball dec}
\end{align}
By introducing $x\equiv \eta/\eta_{\rm{dec}}$, we obtain
\begin{align}
    \frac{{\rm{d}}}{{\rm{d}}x} L_a(T)
    =&
    \frac{12}{5}
    \frac{x(x+2x_*)}{1+3x_*}
    \left(
    1-\frac{x^2(x+3x_*)}{1+3x_*}
    \right)^{-1/5}
    L_a^{\rm{dec}}
    \nonumber \\
    &
    -
    \frac{45}{4\pi^4g_{*,s}(T_{\rm{dec}})}
    \frac{{\rm{d}}t}{{\rm{d}}x}
    \int{{\rm{d}}\epsilon}
    \;
    \epsilon^2 
    \Gamma_{\nu_a}(p,T)
    \left[
        \theta_M^2(\epsilon,T)
        f_{\nu_a}(\epsilon,T)
        \right.
        -
        \left.
       \bar{\theta}_M^2(\epsilon,T)
        f_{\bar{\nu}_a}(\epsilon,T)
    \right]
    \ ,
    \nonumber \\
    \frac{{\rm{d}}}{{\rm{d}}x} 
    f_{\nu_s}(\epsilon, T)
    =&
    \frac{{\rm{d}}t}{{\rm{d}}x}
    \Gamma_{\nu_a}(p,T)
        \left(
        \theta_M^2(\epsilon,T)
        f_{\nu_a}(\epsilon,T)
        +
        \bar{\theta}_M^2(\epsilon,T)
        f_{\bar{\nu}_a}(\epsilon,T)
        \right).
    \label{eq : basic equation}
\end{align}
Here, ${\rm{d}}t/{\rm{d}}x$ is given by
\begin{align}
    \frac{{\rm{d}}t}{{\rm{d}}x}
    =
    a\eta_{\rm{dec}}
    =
    a(\eta_{{\rm{eq}},1})\eta_{\rm{dec}}\left(
    \frac{x^2}{x_*^2}
    +\frac{2x}{x_*}
    \right),
\end{align}
with $x_* \equiv \eta_*/\eta_\mathrm{dec}$.
We numerically solve the above equations by setting $L_a=0$ and $f_{\nu_s}=0$ at $\eta\ll\eta_{\rm{eq,1}}$, and evaluate the abundance of the sterile neutrinos produced during the lepton number injection. 

After the Q-balls have completely decayed, we calculate $\nu_s$ production process in the same way as in Sec.~\ref{subsec : stationary point approximation}, setting the value of $L_a$ and physical temperature $T$ at the completion of Q-ball decay as the initial conditions.

\begin{figure}[t]
\centering
\begin{tabular}{c} 
    \includegraphics[width=0.7\linewidth]{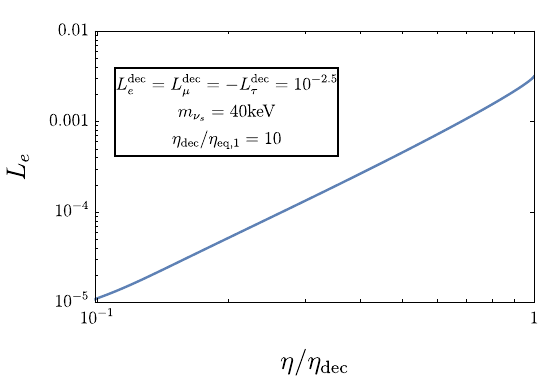}
    \\
    \includegraphics[width=0.7\linewidth]{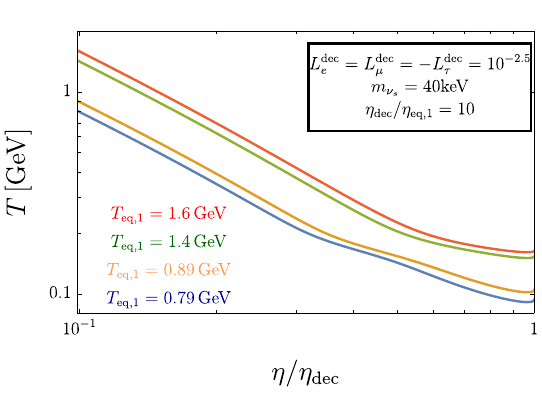}
    \end{tabular}
    \caption{Dependence of $L_e$ (upper panel) and $T$ (lower panel) on the $x=\eta/\eta_{\rm{dec}}$ with $m_{\nu_s}=40$\,keV, $L_e^{\rm{dec}}=10^{-2.5}$ and $\eta_{\rm{dec}}/\eta_{\rm{eq},1}=10$. 
    Different colors correspond to different values of $T_{\rm{eq,1}}$. 
    }
    \label{fig : evolution of L and T}
\end{figure}

\begin{figure}[t]
\centering
     \begin{tabular}{c} 
    \includegraphics[width=0.75\linewidth]{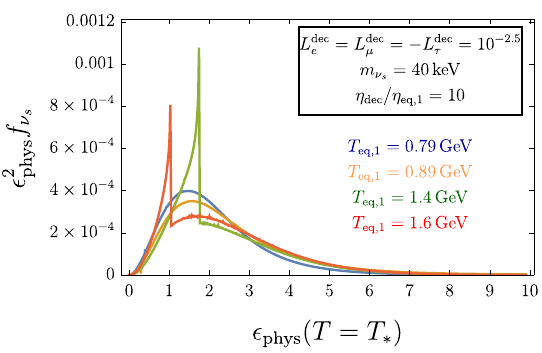}
    \end{tabular}
    \caption{The final spectra of the sterile neutrinos for $m_{\nu_s}=40$\,keV, $L_e^{\rm{dec}}=10^{-2.5}$ and $\eta_{\rm{dec}}/\eta_{\rm{eq},1}=10$.
    Different colors correspond to different values of $T_{\rm{eq,1}}$.}
    \label{fig : spectra during Q ball dec}
\end{figure}

In Fig.~\ref{fig : evolution of L and T}, we show the time evolution of the lepton asymmetry parameter $L_e$ during the Q-ball decay and the cosmic temperature $T$ as functions of $x=\eta/\eta_{\rm{dec}}$.
In Fig.~\ref{fig : spectra during Q ball dec}, we show the final spectra of the sterile neutrinos at $T = T_*$ for various  $T_{\rm{eq,1}}$.
Here, we take $L_e^{\rm{dec}}=L_{\mu}^{\rm{dec}}=-L_\tau^{\rm{dec}}=10^{-2.5}$, $m_{\nu_s}=40$\,keV, and $\eta_{\rm{dec}}/\eta_{\rm{eq},1}=10$. 
If $T_{\rm{eq,1}}$ is smaller than $1$\,GeV ($T_{\rm{dec}}\lesssim0.12$\,GeV), the resonant production takes place before $T=T_{\rm{dec}}$.
After the production, the momenta of sterile neutrinos are redshifted as $\propto a^{-1}$, while the entropy production by Q-balls slows down the decrease of $T$. 
For a smaller $T_{\rm{eq,1}}$ means that entropy production continues for a longer time after $\nu_s$ production, which shifts the peak of the momentum distribution to lower values at $T=T_*$.

\begin{figure}[t]
\centering
\begin{tabular}{c}
    \includegraphics[width=0.7\linewidth]{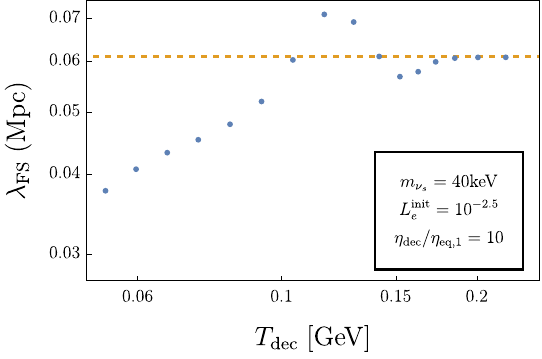}
    \end{tabular}
    \caption{The dependence of the free-streaming length $\lambda_{\rm{FS}}$ on $T_{\rm{dec}}$. 
    Here, we fix $\theta$ to match the DM abundance.
    We also show the free streaming length when resonant production takes place after the lepton asymmetry generation in a dashed horizontal line.}
    \label{fig : FS when nus prod during Qball dec}
\end{figure}

In contrast, for $T_{\rm{eq,1}}$ larger than $1$\,GeV ($T_{\rm{dec}}\gtrsim0.12$\,GeV), high momentum modes are produced after the Q-balls have completely decayed, while low momentum modes are produced during the decay process. 
During the decay, resonant production is enhanced due to the slower temperature decrease due to the entropy production, allowing the resonance condition (Eq.~\eqref{eq : resonance cond for T}) to be satisfied for a longer time. 
However, right after the decay ends, $T$ begins to drop rapidly, and the resonance conditions are satisfied for a shorter time, which is the same situation as in the numerical simulations performed in Sec.~\ref{sec : resonant production}. 
As a result, a sharp peak emerges in the final spectrum at the mode that is resonantly produced right at the end of the Q-ball decay. 
The reason why the spectrum seems discontinuous around the sharp peak is that ${{\rm{d}}t}/{{\rm{d}}T}$ is discontinuous at $t=t_{\rm{dec}}$.
For $1\,\mathrm{GeV} \lesssim T_{\rm{eq},1}\lesssim 1.3$\,GeV ($0.12\,\mathrm{GeV}<T_{\rm{dec}}<0.14$\,GeV), the peak appears at a relatively high momentum mode and thus increases the free-streaming length compared with the case where resonant production occurs after the decay of Q-balls. 
Conversely, $T_{\rm{eq},1}\gtrsim 1.3$\,GeV ($T_{\rm{dec}}>0.14$\,GeV) leads to a peak at low momentum mode, reducing the free-streaming length compared with the case where resonant production occurs after the decay of Q-balls. 
This trend is also seen in Fig.~\ref{fig : FS when nus prod during Qball dec}, where we illustrate how the free-streaming length depends on $T_{\rm{dec}}$. 
The dependence of $\lambda_{\rm{FS}}$ on $T_{\rm{eq},1}$ can be qualitatively understood based on the discussion above. 

\begin{figure}[t]
\centering
\begin{tabular}{c} \includegraphics[width=0.7\linewidth]{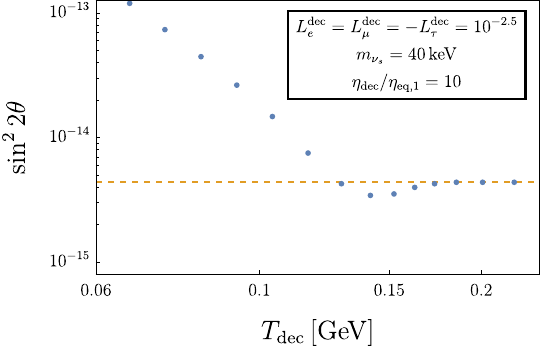}
    \end{tabular}
    \caption{The dependence of $\theta$ to explain all DM on $T_{\rm{dec}}$.
    We also show the required $\theta$ when resonant production takes place after the lepton asymmetry generation in a dashed horizontal line.}
    \label{fig : theta when nus prod during Qball dec}
\end{figure}

In Fig.~\ref{fig : theta when nus prod during Qball dec}, we show the dependence of $\theta$ to explain all DM on $T_{\rm{dec}}$. 
With $T_{{\rm{eq}},1}\lesssim 1.1$\,GeV ($T_{{\rm{dec}}}\lesssim 0.13$\,GeV), the required $\theta$ is larger than the case with $\nu_s$ production after the Q-ball decay because the value of $L_a$ is still significantly smaller at the $\nu_s$ production and the produced $\nu_s$ is diluted by the Q-ball decay.
On the other hand, if $1.1$\,GeV$\lesssim T_{{\rm{eq}},1}\lesssim 1.8$\,GeV ($0.13$\,GeV$\lesssim T_{{\rm{dec}}}\lesssim 0.17$\,GeV), the required $\theta$ can be smaller. 
This is because a large fraction of sterile neutrinos are produced during the process of entropy production by the decay of Q-balls, which prolongs the resonance duration as discussed above.
For larger $T_\mathrm{eq,1}$, this effect becomes less significant, and then, the required $\theta$ becomes independent of $T_\mathrm{eq,1}$ because the Q-balls completely decay before the resonant production of sterile neutrinos.

In summary, it is found that if a fraction of sterile neutrinos is produced right before the Q-ball decay is completed, the observational constraints are relaxed since the required mixing angle becomes smaller, and the final spectrum is peaked at a lower momentum mode.

\section{Summary}
\label{sec : discussion and summary}

Sterile neutrinos can be produced via mixing with the SM neutrinos and are a fascinating candidate for DM.
Given the constraints from the X-ray observations, however, all DM cannot be accounted for in the minimal scenario, i.e., via the DW mechanism.
This constraint is significantly relaxed by resonant production with a large lepton asymmetry.
However, the numerical simulation of the resonant production requires a significantly fine binning of the momentum because the resonance takes place in a narrow and time-dependent range of momenta.
This trend is more significant for a larger lepton asymmetry and a smaller sterile neutrino mass.

In this paper, we refined the numerical scheme to correctly capture the resonant production process of sterile neutrinos as dark matter. 
By adopting a dynamical binning method that always finely resolves the resonant modes, we observed a convergence of the numerical results with a bin number of $N_\mathrm{bin} \gtrsim 100$.
This contrasts with the results of a static linear binning, which does not converge even for $N_\mathrm{bin} = 10^4$.

With this refined numerical method, we evaluated the momentum distributions of sterile neutrinos for various parameter sets.
By scanning the parameter region of sterile neutrinos,
we obtained the constraints from X-ray and Lyman $\alpha$ observations, $L_a\gtrsim \mathcal{O}(10^{-3})$ and $m_{\nu_s}\gtrsim 20$\,keV.
Moreover, we investigated the scenario where the large lepton asymmetry is produced via the Q-ball decay with lepton charge in the framework of Affleck-Dine leptogenesis.
In particular, we studied the case where a fraction of sterile neutrinos is produced during the decay process of Q-balls. 
The entropy production by the decay of Q-balls slows down the redshift of cosmic temperature, and then the resonance duration is prolonged, enhancing the final spectrum, especially if the sterile neutrinos are produced right before the decay process is terminated. 
Furthermore, in such a case, the final spectrum is peaked at a lower momentum than in the case where sterile neutrinos are produced after the injection of lepton asymmetry.
This result implies the possibility that the resonant production of sterile neutrinos during the Q-ball decay relaxes the constraints from the structure formation.

However, in discussing the constraint from the structure formation, we have just compared the averaged free-streaming length of sterile neutrinos with that of thermal relic DM. 
To obtain more robust constraints, we need to perform a more detailed numerical study of the effect on structure formation, by using the final $\nu_s$ spectra obtained in this paper.
Furthermore, in the numerical computation of sterile neutrino production during the injection of lepton number, we approximated that Q-balls have monochromatic charge and mass distributions. 
The non-monochromatic mass distribution can slightly change the time-temperature relation during the lepton number injection, modifying the final spectrum of sterile neutrinos. This issue will require a detailed lattice simulation of Q-ball formation, which we leave for future work.

\begin{acknowledgments}
This work was supported by JSPS KAKENHI Grant Nos. 25K07297(M.K.), 23KJ0088 (K.M.), 24K17039 (K.M.), and 25KJ1164 (K.K.).
K.K. was supported by the Spring GX program.
\end{acknowledgments}

\appendix

\section{Fitting formulae for \texorpdfstring{$y_e$}{}}
\label{append : ye fitting}

\begin{figure}[t]
\centering
\begin{tabular}{c} \includegraphics[width=0.7\linewidth]{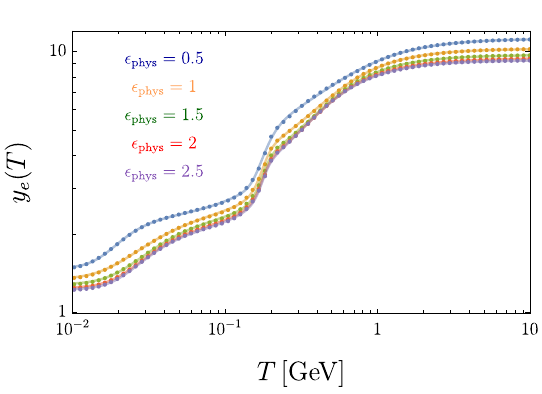}
    \end{tabular}
    \caption{Comparison of fitting formulae and the numerical results obtained in Ref.~\cite{Ghiglieri:2015jua} regarding $y_e$. The blue, orange, green, red, purple lines correspond to $\epsilon_{\rm{phys}}=0.5,1,1.5,2,2.5$, respectively. The dots represent the numerical results while the lines correspond to our fitting formulae.}
    \label{fig : Comparison of fitting formulae for ye}
\end{figure}

Here, we present the fitting formulae for $y_e(\epsilon, T)$ defined in Eq.~\eqref{eq : thermal width}.
Regarding $y_e(\epsilon,T)$, we use the data obtained by Ref.~\cite{Ghiglieri:2015jua}, approximating that both the active and sterile neutrinos are relativistic at the production and neglecting the chemical potential of active neutrinos. Since Ref.~\cite{Ghiglieri:2015jua} presents numerical results for discrete values of $\epsilon_{\rm{phys}}=0.25i$ with $i=1,2,\cdots,40$ for fixed $T$, we define a fitting formula for the discrete $\epsilon_{\rm{phys}}$, given by
\begin{align}
    y_e(\epsilon_{\rm{phys}},T)
    =
    \sum_{i=1}^4
    \exp
    \left(
    a_i
    +
    b_i
    \left(
    1+\tanh
    \left(
    \frac{\log(T)+c_i}{d_i}
    \right)
    \right)
    \right),
\end{align}
where $a_i,b_i,c_i,d_i$ are coefficients defined at each discrete values of $\epsilon_{\rm{phys}}$. 
We fit the parameters $a_i,b_i,c_i,d_i$ with $\epsilon_{\rm{phys}}<2.5$ by using the NonlinearModelFit method in Mathematica. On the other hand, we checked that we can neglect $\epsilon$ dependence with $\epsilon_{\rm{phys}}>2.5$, which enables us to use the numerical result of $\epsilon_{\rm{phys}}=2.5$ for such modes. We present the comparison of numerical results and our fitting formulae in Fig.~\ref{fig : Comparison of fitting formulae for ye}.
Based on this, for a given $T$, we compute $y_e$ by linear interpolation with respect to $\epsilon_{\rm{phys}}$. 

\section{Concrete fitting formulae for the susceptibilities}
\label{append : susceptibilities}
\begin{figure}[t]
\centering
\begin{tabular}{c} \includegraphics[width=0.49\linewidth]{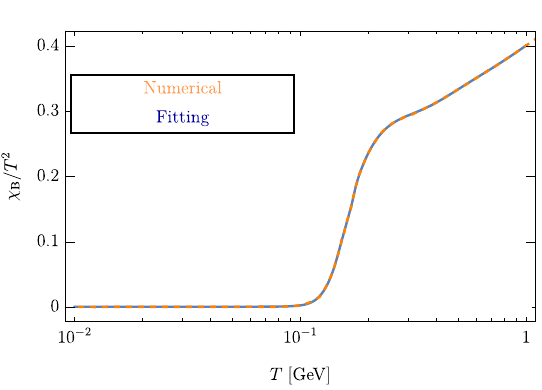}
\includegraphics[width=0.49\linewidth]{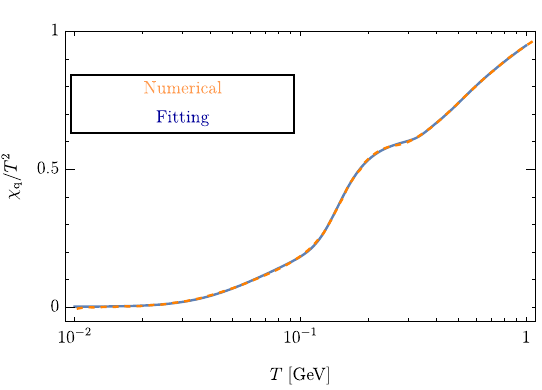}
\\
\includegraphics[width=0.49\linewidth]{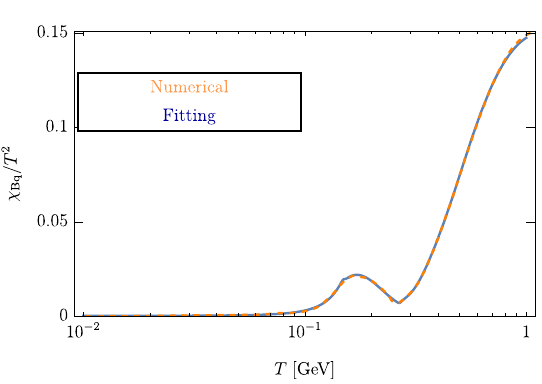}
    \end{tabular}
    \caption{The comparison of fitting formulae (blue solid lines) and numerical results (orange dashed lines) discussed in Ref.~\cite{Venumadhav:2015pla} for $\chi_{B}$ (upper left), $\chi_q$ (upper right), and $\chi_{Bq}$ (bottom).}
    \label{fig : chi B}
\end{figure}

Here, we summarize the fitting formulae for thermodynamic variables $\chi$.
Regarding $\chi$, the integration of the Fermi-Dirac distributions gives~\cite{Ghiglieri:2015jua}
\begin{align}
\chi(m,T)=
\frac{m^2}{\pi^2}
\sum_{n=1}^{\infty}
(-1)^{n+1}K_2
\left(
\frac{nm}{T}
\right),
\end{align}
where $K_2$ is the modified Bessel function. Here, in the numerical computation, we approximate the above by
\begin{align}
\chi(m,T)=
\begin{cases}
   \frac{m^2}{\pi^2}K_2
\left(
\frac{nm}{T}
\right)& (m > 4T) 
\\[0.4em]
   \frac{m^2}{\pi^2}\sum_{n=1}^{10}
(-1)^{n+1}K_2
\left(
\frac{nm}{T}
\right)&(T/10 < m <4T) 
\\[0.4em]
   \frac{1}{6}T^2& (m<T/10)
\end{cases},
\end{align}
Regarding $\chi_q$ and $\chi_B$, we use the following fitting formulae :
\begin{align}
    \frac{\chi_{B/q}}{T^2}=\sum_{i}
    \exp
    \left(
    a_{B/q}^{i}+b_{B/q}^i
    \left(
    1+\tanh
    \left(
    \frac{\log(T/1{\rm{GeV}})+c_{B/q}^i}{d_{B/q}^i}
    \right)
    \right)
    \right),
\end{align}
where the values of $a_{B/q}$, $b_{B/q}$, $c_{B/q}$ and $d_{B/q}$ are given in Tables~\ref{Table : Fitting parameters of chiB} and~\ref{Table : Fitting parameters of chiq}.
\begin{table}[tb]
    \renewcommand{\arraystretch}{1.1}
    \centering
    \begin{tabular}{|c|c|c|c|c|} 
    \hline
    $i$ & $a_B^i$ & $b_B^i$ & $c_B^i$ & $d_B^i$
    \\
    \hline\hline
    $1$ & $-23.4$ & $-26.8$ & $-15.97$ & $-130.0$
    \\
    \hline
    $2$ & $-9.40$ & $3.15$ & $1.23$ & $0.457$
    \\
    \hline
    $3$ & $-9.82$ & $2.68$ & $1.99$ & $0.0829$
    \\
    \hline
    $4$ & $-8.98$ & $3.66$ & $0.996$ & $1.17$
    \\
    \hline
    $5$ & $-8.57$ & $3.63$ & $2.19$ & $0.347$
    \\
    \hline
    $6$ & $-9.77$ & $2.70$ & $1.79$ & $0.0368$
    \\
    \hline
    \end{tabular}
    \caption{Fitting parameters for $\chi_B$.}
    \label{Table : Fitting parameters of chiB}
\end{table}
\begin{table}[tb]
    \renewcommand{\arraystretch}{1.1}
    \centering
    \begin{tabular}{|c|c|c|c|c|} 
    \hline
    $i$ & $a_q^i$ & $b_q^i$ & $c_q^i$ & $d_q^i$
    \\
    \hline\hline
    $1$ & $0.727$ & $-0.478$ & $4.08$ & $-0.396$
    \\
    \hline
    \end{tabular}
    \caption{Fitting parameters for $\chi_q$.}
    \label{Table : Fitting parameters of chiq}
\end{table}
On the other hand, as for $\chi_{Bq}$, we cannot use the above form in $0.15\,{\rm{GeV}}<T<0.27\,{\rm{GeV}}$. Thus, we adopt the following formula :
\begin{align}
    \frac{\chi_{Bq}}{T^2}
    &=\frac{1}{2}\left(
    1+\tanh
    \left(
    50\log
    \left(
    \frac{0.15\,{\rm{GeV}}}{T}\right)\right)\right)\chi_{Bq,\rm{low}}
    \nonumber \\
    &+\frac{1}{4}\left(
    1+\tanh
    \left(
    50\log
    \left(
    \frac{T}{0.15\,{\rm{GeV}}}\right)\right)\right)
    \left(
    1+\tanh
    \left(
    50\log
    \left(
    \frac{0.27\,{\rm{GeV}}}{T}\right)\right)\right)\chi_{Bq,\rm{mid}}
    \nonumber \\
    &+\frac{1}{2}\left(
    1+\tanh
    \left(
    50\log
    \left(
    \frac{T}{0.27\,{\rm{GeV}}}\right)\right)\right)\chi_{Bq,\rm{high}}.
\end{align}
Here, $\chi_{Bq,\rm{low}}$ and $\chi_{Bq,\rm{high}}$ are given by
\begin{align}
    \chi_{Bq,\rm{high/low}}
    =\sum_{i}
    \exp
    \left(
    a_{Bq,\rm{high/low}}^{i}+b_{Bq,{\rm{high/low}}}^i
    \left(
    1+\tanh
    \left(
    \frac{\log(T/1{\rm{GeV}})+c_{Bq,{\rm{high/low}}}^i}{d_{Bq,{\rm{high/low}}}^i}
    \right)
    \right)
    \right).
     \label{eq : chi BQ high and chi BQ low}
\end{align}
Regarding $\chi_{Bq,\rm{mid}}$, the above fitting form cannot be a good approximation. Alternatively, we use the following Fermi-Dirac approximation with the numerical correction as
\begin{align}
    \chi_{Bq,{\rm{mid}}}
    =
    \frac{1}{3}
    \left(
    4(\chi(m_u,T) + \chi(m_c,T))
    -2(\chi(m_d,T)
    +\chi(m_s,T)
    +\chi(m_b,T))
    \right)
    \times\chi_{Bq,{\rm{mid,num}}},
\end{align}
where $m_u$, $m_d$, $m_s$, $m_c$, $m_t$, and $m_b$ are the masses of up, down, strange, charm, top, and bottom quarks. $\chi_{{\rm{Bq,mid,num}}}$ is the numerical correction given by
\begin{align}
    \chi_{Bq,{\rm{mid,num}}}
    =
    a_{Bq}
    \log\left(
    \frac{T}{1\,{\rm{GeV}}}
    \right)
    +
    b_{Bq}
    \left(
    \log
    \frac{T}{1\,{\rm{GeV}}}
    \right)^2
    +
    c_{Bq}
    \left(
    \log
    \frac{T}{1\,{\rm{GeV}}}
    \right)^3
    +
    d_{Bq}
    \left(
    \log
    \frac{T}{1\,{\rm{GeV}}}
    \right)^4.
    \label{eq : chi BQ mid}
\end{align}
The values of coefficients in Eqs.~\eqref{eq : chi BQ high and chi BQ low} and~\eqref{eq : chi BQ mid} are given in Tables~\ref{Table : Fitting parameters of chiBq high}, \ref{Table : Fitting parameters of chiBq low}, and \ref{Table : Fitting parameters of chiBq mid}.
We show the numerical results obtained in Ref.~\cite{Venumadhav:2015pla} and our fitting formulae in Fig.~\ref{fig : chi B}.
\begin{table}[tb]
    \renewcommand{\arraystretch}{1.1}
    \centering
    \begin{tabular}{|c|c|c|c|c|} 
    \hline
    $i$ & $a_{Bq,{\rm{high}}}^i$ & $b_{Bq,{\rm{high}}}^i$ & $c_{Bq,{\rm{high}}}^i$ & $d_{Bq,{\rm{high}}}^i$
    \\
    \hline\hline
    $1$ & $-12.6$ & $29.4$ & $-837$ & $114$
    \\
    \hline
    $2$ & $-6.73$ & $-23.7$ & $235$ & $-227$
    \\
    \hline
    $3$ & $5.28$ & $-57.7$ & $104$ & $-78.9$
    \\
    \hline
    $4$ & $-7.37$ & $-22.5$ & $238$ & $-222$
     \\
    \hline
    \end{tabular}
    \caption{Fitting parameters for $\chi_{Bq,{\rm{high}}}$.}
    \label{Table : Fitting parameters of chiBq high}
\end{table}
\begin{table}[tb]
    \renewcommand{\arraystretch}{1.1}
    \centering
    \begin{tabular}{|c|c|c|c|c|} 
    \hline
    $i$ & $a_{Bq,{\rm{low}}}^i$ & $b_{Bq,{\rm{low}}}^i$ & $c_{Bq,{\rm{low}}}^i$ & $d_{Bq,{\rm{low}}}^i$
    \\
    \hline\hline
    $1$ & $-18.0$ & $-27.5$ & $-51.1$ & $-84.9$
    \\
    \hline
    $2$ & $-37.5$ & $17.4$ & $-0.0216$ & $0.336$
    \\
    \hline
    $3$ & $-37.5$ & $17.4$ & $0.368$ & $0.331$
    \\
    \hline
    $4$ & $-8.80$ & $11.7$ & $0.9998$ & $1.35$
    \\
    \hline
    $5$ & $-37.5$ & $17.4$ & $0.196$ & $0.333$
    \\
    \hline
    \end{tabular}
    \caption{Fitting parameters for $\chi_{Bq,{\rm{low}}}$.}
    \label{Table : Fitting parameters of chiBq low}
\end{table}
\begin{table}[tb]
    \renewcommand{\arraystretch}{1.1}
    \centering
    \begin{tabular}{|c|c|c|c|} 
    \hline
    $a_{Bq,{\rm{mid}}}$ & $b_{Bq,{\rm{mid}}}$ & $c_{Bq,{\rm{mid}}}$ & $d_{Bq,{\rm{low}}}$
    \\
    \hline\hline
    $0.664$ & $1.397$ & $0.965$ & $0.215$
    \\
    \hline
    \end{tabular}
    \caption{Fitting parameters for $\chi_{Bq,{\rm{mid,num}}}$.}
    \label{Table : Fitting parameters of chiBq mid}
\end{table}

\section{Relation between \texorpdfstring{$L_a$}{} and \texorpdfstring{$V_a$}{}}
\label{append : relation between La and Va}

Here, we present the relation between $L_a$ and $\mathcal{L}_a$, which appears in $V_a$.
By solving Eqs.~\eqref{eq : succeptibilities}, we obtain
\begin{align}
\mathcal{L}_a
&=\frac{\chi_0(2\mu_a+\sum_{b\neq a}\mu_b)+
2(2\sin^22\theta_w+\frac{1}{2})\chi_a(\mu_a-\mu_q)
+2(2\sin^22\theta_w-\frac{1}{2})\sum_{b\neq a}\chi_b(\mu_b-\mu_q)}{s_{\rm{tot}}},
\label{eq : relation between La and chi}
\end{align}
where $\chi_a\equiv \chi(m_{e_a},T)$, $\chi_{\rm{tot}}\equiv \chi_e+\chi_\mu+\chi_\tau$, $\chi_0\equiv\chi(0,T)$, and
\begin{align}
\mu_q &= 
-2s_{\rm{tot}}
\chi_B
\left[
4\chi_e\chi_\mu\chi_\tau(L_e+L_\mu+L_\tau)
+\chi_0^2\sum_a \chi_aL_a
+2\chi_0\sum_a \frac{\chi_e\chi_\mu\chi_\tau}{\chi_a}(L_e+L_\mu+L_\tau-L_a)
\right]
\nonumber \\
&\times
\left[
\chi_0^3
\left(\chi_{BQ}^2-\chi_B(2\chi_{\rm{tot}}+\chi_Q)
\right)
+
2\chi_0^2
\left(
\chi_{BQ}^2\chi_{\rm{tot}}
-\chi_B
\left(
\chi_Q\chi_{\rm{tot}} + 4(\chi_e\chi_\mu+\chi_\mu\chi_\tau+\chi_\tau\chi_e)
\right)
\right)
\right.
\nonumber \\
&
+4\chi_0
\left(
(\chi_{BQ}^2
-\chi_B\chi_Q)(\chi_e\chi_\mu+\chi_\mu\chi_\tau+\chi_\tau\chi_e)
-6\chi_B\chi_e\chi_\mu\chi_\tau
\right)
+8\chi_e\chi_\mu\chi_\tau(\chi_{BQ}^2-\chi_B\chi_Q)
\Bigr]^{-1},
\nonumber \\
\mu_a
&=
\frac{L_as_{\rm{tot}}+2\chi_a\mu_q}{\chi_0+2\chi_a},
\nonumber \\
\mu_{e_a}
&=\mu_a-\mu_q.
\label{eq : muq mua and muea}
\end{align}
On the other hand, in Sec.~\ref{Subsec : resonance during lepton number generation}, we replace as $s_{\rm{tot}}\to s_{\rm{tot}}(T=T_{\rm{dec}})(a_{\rm{dec}}/a)^3$ in Eqs.~\eqref{eq : relation between La and chi} and~\eqref{eq : muq mua and muea}.

\section{Concrete expressions of bin transformations}
\label{Append : beta and vres}

Here, we present the expression of the functions used for the dynamical binning.
Concrete forms of $\beta$ and $v_{\rm{res}}$, which are defined in Eq.~\eqref{eq : relation between u and v}, are given by
\begin{align}
    \beta(T)
    &=
    \left[
    2^{2/3}\cdot 3^{1/3}
    (-5\alpha^2-8\alpha^3+4\alpha^4)-3u_{\rm{res}}^2(3+8\alpha+4\alpha^2)-\sqrt{3}F_1(\alpha,u_{\rm{res}})
    \right.
    \nonumber \\
    &+u_{\rm{res}}
    \left(
9+2\sqrt{3}F_1(\alpha,u_{\rm{res}})
    +4\alpha^2(3-2^{2/3}\cdot3^{1/3}F_2(\alpha,u_{\rm{res}}))
    +4\alpha(6-2^{-1/3}\cdot3^{1/3}F_2(\alpha,u_{\rm{res}}))
    \right)
    \nonumber \\
    &
    \left.
    +2^{4/3}\cdot3^{2/3}\alpha(1+2\alpha) F_2(\alpha,u_{\rm{res}})+2^{4/3}\cdot3^{1/6}F_1(\alpha,u_{\rm{res}})F_2(\alpha,u_{\rm{res}})
    +
    (2-8\alpha^2)F_2(\alpha,u_{\rm{res}})^2
    \right]
    \nonumber \\
    &/(2(1+2\alpha)F_2(\alpha,u_{\rm{res}})^2),
    \nonumber \\
    v_{\rm{res}}(T)
    &=
    -
    \frac{
    -18(u_{\rm{res}}+\alpha)
    +
    2^{4/3}
    \cdot
    3^{5/3}
    F_2(\alpha,u_{\rm{res}})^{-1}
    (-3u_{\rm{res}}+3u_{\rm{res}}^2
    +(\alpha-1)\alpha)
    +2^{2/3}\cdot
    3^{4/3}
    F_2(\alpha,u_{\rm{res}})
    }{18(1+2\alpha)},
    \label{eq : beta and v res}
\end{align}
where 
\begin{align}
    F_1 (\alpha,u_{\rm{res}})
    &\equiv (1+2\alpha)\sqrt{-(54u_{\rm{res}}^3-27u_{\rm{res}}^4+18\alpha^2 u_{\rm{res}}+\alpha^3(\alpha-4)-9u_{\rm{res}}^2(3+2\alpha^2))},
    \nonumber \\
    F_2(\alpha,u_{\rm{res}}) &\equiv \left(
    27u_{\rm{res}}^2-18u_{\rm{res}}^3+9\alpha^2-9u_{\rm{res}}(1+2\alpha^2)+\sqrt{3}F_1(\alpha,u_{\rm{res}})
    \right)^{1/3}.
\end{align}
Here, we have imposed the condition that $u=0,1$ exactly correspond to $v=0,1$.

\section{Analysis with full quantum kinetic equations (QKEs)}
\label{append : full QKE}
Now let us directly check the validity of the stationary point approximation. 
To this end, we simultaneously solve Eqs.~\eqref{eq : first line in density operator equation},~\eqref{eq : second line in density operator equation}, and~\eqref{eq : time evolution of L}, without using the stationary point approximation. In this appendix, we neglect the lepton number redistribution to the charged lepton sector just for simplicity, but the conclusion will not be changed if we take it into account.
We choose $m_{\nu_s}=10\,$keV and $L_a^{\rm{init}}=10^{-3}, \ 5\times 10^{-3}$ as benchmark parameters, for which the values of $D_{\rm{stat}}$ are relatively small $\sim \mathcal{O}(0.1\,\text{--}\,1)$, and hence the stationary point approximation becomes non-trivial.

Since we take $(v,T)$ instead of $(\epsilon, T)$ as the basic variables, the time evolution of density matrices is given by
\begin{align}
\left.
    \frac{\partial}{\partial T}
    \rho(T,\epsilon(v,T))
    \right|_{v}
    =
    \left.
    \frac{\partial }{\partial T}
    \rho(T,\epsilon)
    \right|_{\epsilon}
    +
    \left(
    \frac{\left.\frac{\partial \epsilon}{\partial T}
    \right|_v}
    {\left.\frac{\partial \epsilon}{\partial v}
    \right|_T}
    \right)
    \frac{\partial}{\partial v}
    \rho(T,\epsilon).
\end{align}
Here, regarding the first term in the right hand side, we use Eqs.~\eqref{eq : first line in density operator equation} and~\eqref{eq : second line in density operator equation}. 
As for the coefficient in the second term, by using Eqs.~\eqref{eq : def of u},~\eqref{eq : relation between u and v} and~\eqref{eq : def of u res}, we obtain the numerator and the denominator as
\begin{align}
    \left.\frac{\partial \epsilon}{\partial T}
    \right|_v
    &=
    \frac{{\rm{d}}\epsilon_{\rm{low}}(T)}{{\rm{d}}T}
    \nonumber \\
    &~+
    \left(
    \epsilon_{\rm{max}}-\epsilon_{\rm{min}}
    \right)
    \left[
    -\left(\alpha+3\beta(T)(v-v_{\rm{res}}(T))^2
    \right)
    \frac{{\rm{d}}v_{\rm{res}}(T)}{{\rm{d}}T}
    +
    (v-v_{\rm{res}}(T))^3
    \frac{{\rm{d}}\beta(T)}{{\rm{d}}T}
    \right],
    \\
    \left.\frac{\partial \epsilon}{\partial v}
    \right|_T
    &=
    (\epsilon_{\rm{max}}-\epsilon_{\rm{min}})
    \left(
    \alpha+3\beta(v-v_{\rm{res}}(T))^2
    \right),
\end{align}
where ${\rm{d}}\epsilon_{\rm{low}}/{\rm{d}}T$, ${\rm{d}}v_{\rm{res}}/{\rm{d}}T$ and ${\rm{d}}\beta/{\rm{d}}T$ are calculated from Eqs.~\eqref{eq : epsilon low} and~\eqref{eq : beta and v res}, respectively.
We also solve simultaneously
\begin{align}
\frac{{\rm{d}}}{{\rm{d}}t}
L_{a}
=
-\frac{45}{4\pi^4g_{*,s}(T_i)}
m_{\nu_s}^2\sin2\theta\int{\rm{d}}v
    \left.
    \frac{\partial\epsilon}{\partial v}
    \right|_{T}
    \epsilon^2 
\;{\rm{Im}}(\rho_{as}-\bar{\rho}_{as}).
\end{align}

In numerical computation, to correctly treat the first term, we take the timestep $\Delta T$ such that $|\Delta T| \ll \Gamma_{\nu_a}^{-1}\cdot{\rm{d}}T/{\rm{d}}t$ for all modes under resonance, to avoid the artificial overdamping of the numerical solution to the stationary solution. Here, the criteria for the resonance is chosen such that
\begin{align}
    \left|V_a -\frac{m_{\nu_s}^2}{2p}\cos2\theta
    \right|
    <
    K_{\rm{res}}\left|\frac{\Gamma_{\nu_a}}{2}
    \right|,
\end{align}
where we choose $K_{\rm{res}}=10$.
Furthermore, we treat the second term using the upwind method such that
\begin{align}
\frac{\partial}{\partial v}
    \rho(T,\epsilon)
    &\simeq 
    \frac{\rho(T,\epsilon(v_{i+1},T))
    -\rho(T,\epsilon(v_{i},T))}{\Delta v}
    \;\;{\rm{for\;}}{\left.\frac{\partial \epsilon}{\partial T}
    \right|_v}
    \cdot{\left.\frac{\partial \epsilon}{\partial v}
    \right|_T}>0
    \nonumber \\
    &\simeq 
    \frac{\rho(T,\epsilon(v_{i},T))
    -\rho(T,\epsilon(v_{i-1},T))}{\Delta v}
    \;\;{\rm{for\;}}{\left.\frac{\partial \epsilon}{\partial T}
    \right|_v}
    \cdot{\left.\frac{\partial \epsilon}{\partial v}
    \right|_T}<0
\end{align}
and take the timestep to satisfy the CFL condition
\begin{align}
    \Delta T < 
    \left(
    \frac
{\left.\frac{\partial \epsilon}{\partial v}
    \right|_T}
    {\left.\frac{\partial \epsilon}{\partial T}
    \right|_v}
    \right)
    \Delta v.
\end{align}
By considering the above, we choose the timestep by
\begin{align}
    |\Delta T|
    =
    {\rm{min}}
    \left[
    C_{\rm{CFL}}
    \cdot
    \min\limits_{v}
    \left|
    \frac
{\left.\frac{\partial \epsilon}{\partial v}
    \right|_T}
    {\left.\frac{\partial \epsilon}{\partial T}
    \right|_v}
    \right|
    \Delta v
    ,
\;C_{\rm{res}}
\cdot
\left| 
    \frac{{\rm{d}}T}{{\rm{d}}t}
\right|
\min_{v \;:\; \text{under res}}
\Gamma_{\nu_a}\,\bigl(\epsilon(v,T),\,T\bigr)^{-1}
    \right],
\end{align}
where we take $C_{\rm{CFL}}=0.5$ and $C_{\rm{res}}=0.1$.

\begin{figure}[t]
\centering
\begin{tabular}{c} 
    \includegraphics[width=0.75\linewidth]{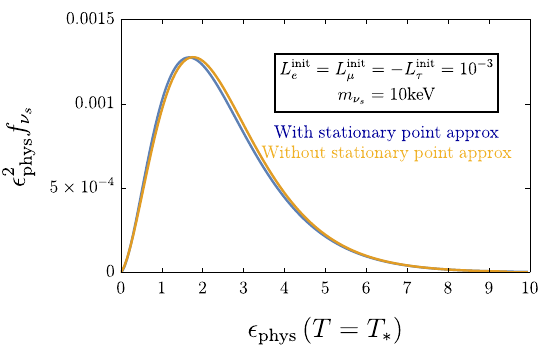}
    \\
    \includegraphics[width=0.75\linewidth]{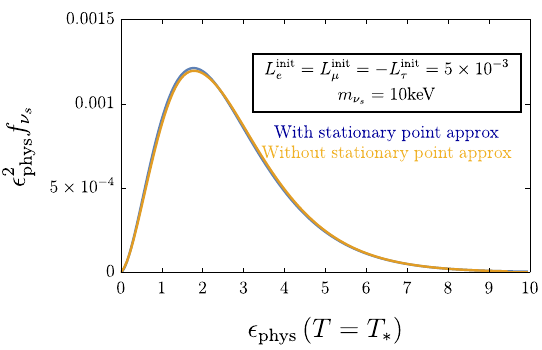}
    \end{tabular}
    \caption{Comparison of final spectra between the computation within stationary point approximation and without stationary point approximation for $m_{\nu_s}=10$~keV and $L_e^{\rm{init}}=L_{\mu}^{\rm{init}}=-L_\tau^{\rm{init}} = 10^{-3}$ (upper) and $5\times 10^{-3}$ (lower).
    }
    \label{fig : comparison with beyond stationary point approx}
\end{figure}

In Fig.~\ref{fig : comparison with beyond stationary point approx}, we show the final sterile neutrino spectra without (orange) and with (blue) the stationary point approximation for the two parameter sets.
It is seen that the differences is only within $\mathcal{O}(1)\%$ in both cases although $D_{\rm{stat}}\lesssim\mathcal{O}(1)$.

\small
\bibliographystyle{JHEP}
\bibliography{Ref}

\end{document}